\documentclass[english]{article}
\usepackage[T1]{fontenc}
\usepackage[latin9]{inputenc}
\usepackage{geometry}
\usepackage{bm}
\usepackage{amsmath}
\usepackage{amssymb}
\usepackage{graphicx}
\usepackage{esint}
\usepackage{babel}
\begin{document}
\global\long\def\sinc{{\rm sinc}}%
\global\long\def\Res{{\rm Res}}%

\title{Calculation of the electromagnetic scattering by non-spherical particles
based on the volume integral equation in the spherical wave function
basis}
\author{Alexey A. Shcherbakov\\ITMO University, Saint-Petersburg, Russia}
\date{}
\maketitle
\begin{abstract}
The paper presents a method for calculation of non-spherical particle
T-matrices based on the volume integral equation and the spherical
vector wave function basis, and relies on the Generalized Source Method
rationale. The developed method appears to be close to the invariant
imbedding approach, and the derivations aims at intuitive demonstration
of the calculation scheme. In parallel calculation of single columns
of T-matrix is considered in detail, and it is shown that this way
not only has a promising potential of parallelization but also yields
an almost zero power balance for purely dielectric particles.
\end{abstract}

\section{Introduction}

An accurate simulation of the light scattering by non-spherical particles
is important for a variety of applications ranging from planetary
science to nanoscale power transfer. A number of methods for achieving
this goal were developed \cite{Mishchenko2000}. When looking from
the point of view of ``arbitrariness'' of possible particle shapes
and a range of treatable size parameters the Discreet Dipole Approximation
(DDA) and the other Volume Integral Equation (VIE) methods appear
to be the most handy while retaining a relative formulation simplicity
\cite{Yurkin2007,Yurkin2018}. The methods are based on the volume
integral equation solution to the three-dimensional Helmholtz equation,
and a three-dimensional equidistant spatial discretization makes it
possible to greatly benefit from fast-Fourier transform based numerical
algorithms. That said, the named methods conventionally lack of another
property being quite important for applications. They yield and output
in form of a response vector given an excitation field vector, while
a complete T-matrix \cite{Mishchenko1996,Mishchenko2004} is often
of interest. The reason is in the mismatch in the field representation,
which is the point-wise storage of the spatial field components in
case of the DDA/VIE, while the T-matrix is conventionally defined
in terms of the spherical vector wave field decomposition.

A possible way to couple the pros of the VIE methods with a direct
T-matrix output is to consider the volume integral equation in the
spherical vector wave function basis together with a spatial discretization
into a set of thin spherical shells instead of small cubic volumes.
This approach was considered in \cite{Johnson1988} within the rationale
of the invariant imbedding technique, and further applied in \cite{Bi2013-1,Bi2013-2,Bi2014}
for analysis of atmospheric ice particle light scattering features.
The equivalence between the invariant imbedding procedure and the
integral-matrix approaches was outlined in \cite{Wriedt2018,Doicu2018}.
The method represents an interesting alternative to the previously
well-developed volume integral methods. This work proposes a different
view on the mentioned approach. In addition within the same rationale
a numerical approach to calculate single columns of the T-matrix is
presented in detail, which was only mentioned in \cite{Johnson1988}.
The latter approach not only possess a potential for parallelization,
but also is shown to yield results which meet the energy conservation
at very high precision.

In order to trace analogies between the methods in spherical and planar
geometries the derivations of this paper are based on the rationale
of the Generalized Source Method (GSM) \cite{Tishchenko2000}, which
was applied previously to the grating diffraction problems \cite{Shcherbakov2012}.
Besides, this logic is chosen to support a further introduction of
the generalized metric sources in the spherical vector wave basis
in analogy with \cite{Shcherbakov2013,Shcherbakov2017}, to be described
in a next paper. The GSM relies on basis solutions which provide a
rigorous way to calculate an output of an arbitrary source current.
Here a homogeneous space basis solution (can be read as the homogeneous
space Green's function) is used to develop a scattering matrix algorithm
being a counterpart of the invariant imbedding method. Then, a basis
solution in a homogeneous spherical layer is used to formulate a scattering
vector algorithm yielding single columns of T-matrices \cite{Shcherbakov2015}.
The terms scattering matrix and scattering vector are defined and
discussed below. Finally, numerical examples are presented demonstrating
an accuracy and convergence rates of the algorithms.

\section{Generalized source method}

The scattering problem being addressed in this work is schematically
demonstrated in Fig. 1. Given a homogeneous scattering particle occupying
a closed three-dimensional volume $\Omega_{p}$ and an external time-harmonic
electromagnetic field with amplitudes ${\bf E}^{ext}\left(\bm{r}\right)$,
${\bf H}^{ext}\left(\bm{r}\right)$ and frequency $\omega$ excited
by some sources ${\bf J}^{ext}\left(\bm{r}\right)$ located outside
the spherical domain $\Omega_{R}^{out}=\left\{ \bm{r}=\left(r,\theta,\varphi\right):\thinspace r\leq R\right\} \ni\Omega_{p}$
containing the particle, here $\left(r,\theta,\varphi\right)$ are
spherical coordinates, one aims at calculation of the total electromagnetic
field being a solution of the time-harmonic Maxwell's equations. This
scattering problem is characterized by a spatially inhomogeneous dielectric
permittivity $\varepsilon\left(\bm{r}\right)$ which equals to some
constant $\varepsilon_{p}$ (possibly complex) inside the volume $\Omega_{p}$,
and to real constant $\varepsilon_{s}$ in the surrounding non-absorbing
medium $\bm{r}\in\mathbb{R}^{3}\backslash\Omega_{p}$. For simplicity,
and aiming at optical applications, within this paper the permeability
is considered to be equal to the vacuum permeability $\mu_{0}$.

\begin{figure}[h]
\begin{centering}
\includegraphics[scale=0.6]{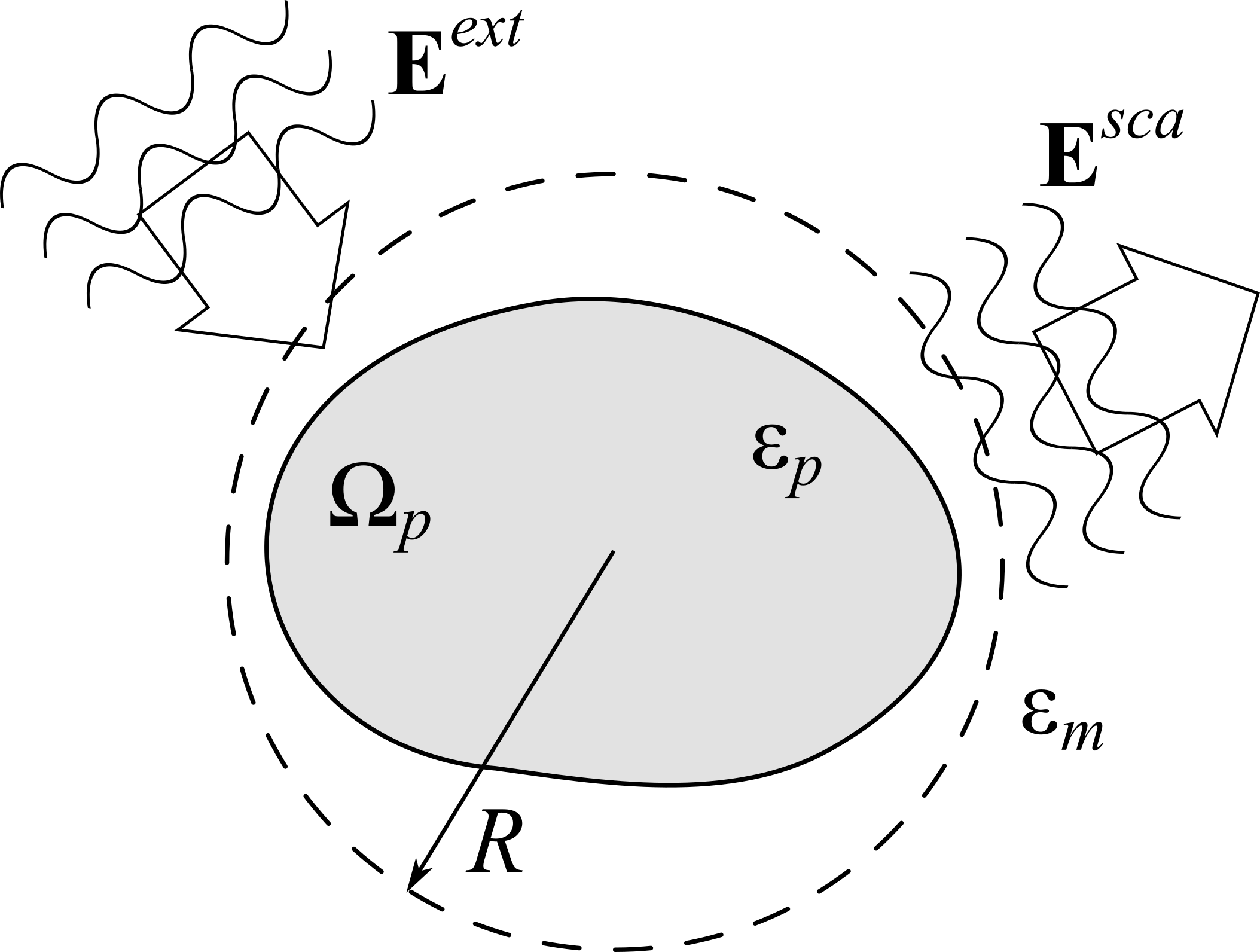}
\par\end{centering}
\caption{Scattering problem being addressed in this work.}

\end{figure}

Assume that solution of the electromagnetic scattering problem is
known for some basis structure characterized by the function $\varepsilon_{b}\left(\bm{r}\right)$
whatever the sources are, and this solution is given by the linear
operator $\mathfrak{S}_{b}$ as follows:
\begin{equation}
{\bf E}={\bf E}^{ext}+\mathfrak{S}_{b}\left({\bf J}\left(\bm{r}\right)\right)\label{eq:gsm_basis}
\end{equation}
Then, in order to find a solution of the initial scattering problem
one has to consider a difference between the initial and the basis
media which gives rise to the generalized source ${\bf J}_{gen}\left(\bm{r}\right)$,
so that the desired self-consistent field ${\bf E}\left(\bm{r}\right)$
appears to meet the following equation
\begin{equation}
{\bf E}={\bf E}^{ext}+\mathfrak{S}_{b}\left({\bf J}_{gen}\left(\bm{r}\right)\right)\label{gsm_solution}
\end{equation}
Conventionally within the volume integral equation methods one takes
${\bf J}_{gen}\left(\bm{r}\right)=-i\omega\left[\varepsilon\left(\bm{r}\right)-\varepsilon_{b}\left(\bm{r}\right)\right]{\bf E}\left(\bm{r}\right)$
\cite{Yurkin2018}. In this work Eq. (\ref{gsm_solution}) is enclosed
in a similar way.

\section{Basis solution}

The declared intention to operate with spherical vector wave function
decompositions generally restricts the choice of the basis medium
to be a set of homogeneous space regions of constant permittivity
separated by concentric spherical interfaces. This is due to the fact
that the reflection and transmission are described in form of diagonal
operators for these basis functions. The basis operator $\mathfrak{S}_{b}$
is explicitly defined by the corresponding Green's function, (e.g.,
\cite{Li1994}). Yet, for the methods developed in this work it is
sufficient to consider only two cases -- a homogeneous isotropic
space, and a single spherical layer bounded by two interfaces. This
section outlines the basis solution in a homogeneous basis space.

In view of the said assume here $\varepsilon_{b}\left(\bm{r}\right)$
to be constant everywhere. Time-harmonic Maxwell's equations in the
basis medium with extracted factor $\exp\left(-i\omega t\right)$

\begin{equation}
\begin{array}{c}
\nabla\times{\bf E}\left(\bm{r}\right)-i\omega\mu_{0}{\bf H}\left(\bm{r}\right)=0,\\
\nabla\times{\bf H}\left(\bm{r}\right)+i\omega\varepsilon_{b}{\bf E}\left(\bm{r}\right)={\bf J}\left(\bm{r}\right)
\end{array}\label{eq:maxwell}
\end{equation}
yield the Helmholtz equation
\begin{equation}
\nabla\times\nabla\times{\bf E}\left(\bm{r}\right)-k_{b}^{2}{\bf E}\left(\bm{r}\right)=i\omega\mu_{0}{\bf J}\left(\bm{r}\right)\label{helmholtz}
\end{equation}
where the wavenumber $k_{b}=\omega\sqrt{\varepsilon_{b}\mu_{0}}$.
Given the spherical coordinates $\left(r,\theta,\varphi\right)$ with
unit vectors $\hat{\bm{e}}_{r}$, $\hat{\bm{e}}_{\theta}$, and $\hat{\bm{e}}_{\varphi}$,
and the modified basis $\left(\hat{\bm{e}}_{r},\hat{\bm{e}}_{+},\hat{\bm{e}}_{-}\right)$
which is related to the spherical one as $\hat{\bm{e}}_{\pm}=\left(\hat{\bm{e}}_{\theta}\pm i\hat{\bm{e}}_{\varphi}\right)/\sqrt{2}$,
the eigen solutions of the homogeneous Helmholtz equation are the
two sets of spherical vector wave functions (see, e.g., \cite{Doicu2006}):
\begin{equation}
\mathcal{{\bf \mathcal{M}}}_{nm}^{1,3}\left(k_{b}\bm{r}\right)=i\frac{\sqrt{2n+1}}{2\sqrt{2\pi}}z_{n}^{1,3}\left(k_{b}r\right)\left[\hat{\bm{e}}_{+}d_{m,1}^{n}\left(\theta\right)+\hat{\bm{e}}_{-}d_{m,-1}^{n}\left(\theta\right)\right]\exp\left(im\varphi\right)\label{eq:svwf_M}
\end{equation}
\begin{align}
\mathcal{{\bf \mathcal{N}}}_{nm}^{1,3}\left(k_{b}\bm{r}\right) & =\frac{\sqrt{n\left(n+1\right)}}{\sqrt{2\pi}}\frac{z_{n}^{1,3}\left(k_{b}r\right)}{k_{b}r}P_{n}^{m}\left(\theta\right)\exp\left(im\varphi\right)\hat{\bm{e}}_{r}\nonumber \\
 & +\frac{\sqrt{2n+1}}{2\sqrt{2\pi}}\frac{\tilde{z}{}_{n}^{1,3}\left(k_{b}r\right)}{k_{b}r}\left[\hat{\bm{e}}_{+}d_{m,1}^{n}\left(\theta\right)-\hat{\bm{e}}_{-}d_{m,-1}^{n}\left(\theta\right)\right]\exp\left(im\varphi\right)\label{eq:svwf_N}
\end{align}
Here integer indices $n,m$ are subject to constraints $n\geq0$,
$\left|m\right|\leq n$. $P_{n}^{m}\left(\theta\right)$ are normalized
associated Legendre polynomials, $d_{m,\pm1}^{n}\left(\theta\right)$
are elements of rotation Wigner $d$-matrices \cite{Varshalovich1988}
with $P_{n}^{m}\left(\theta\right)=\sqrt{\left(2n+1\right)/2}d_{m,0}^{n}\left(\theta\right)$.
These functions of the polar angle obey the orthogonality condition,
which is read via the Kronecker delta-symbol $\delta_{nm}$:
\begin{equation}
\intop_{0}^{\pi}d_{mq}^{n}\left(\theta\right)d_{mq}^{p}\left(\theta\right)\sin\theta d\theta=\frac{2}{2n+1}\delta_{np}.\label{eq:d_orthog}
\end{equation}
Superscripts ''1,3'' correspond to regular and radiating wave functions
respectively, so that $z_{n}^{1}\equiv j_{n}$ is the regular spherical
Bessel function, and $z_{n}^{3}\equiv h_{n}^{(1)}$ is the spherical
Henkel function of the first kind. Besides, $\tilde{z}_{n}\left(x\right)=d\left[xz_{n}\left(x\right)\right]/dx$.
The rotor operator transforms functions (\ref{eq:svwf_M}), (\ref{eq:svwf_N})
one into another:
\begin{equation}
\begin{array}{c}
\nabla\times\mathcal{{\bf \mathcal{M}}}_{nm}^{1,3}\left(k_{b}\bm{r}\right)=k_{b}\mathcal{{\bf \mathcal{N}}}_{nm}^{1,3}\left(k_{b}\bm{r}\right)\\
\nabla\times\mathcal{{\bf \mathcal{N}}}_{nm}^{1,3}\left(k_{b}\bm{r}\right)=k_{b}\mathcal{{\bf \mathcal{M}}}_{nm}^{1,3}\left(k_{b}\bm{r}\right)
\end{array}\label{eq:MN_NM}
\end{equation}
Note that once the center of the spherical coordinate system is fixed,
all spherical harmonic decompositions are made relative to this center,
and no translations are used in this work.

Solution to the Helmholtz equation (\ref{helmholtz}) in the volume
integral form is written via the free-space dyadic Green's function
${\bf G}\left(\bm{r}-\bm{r}'\right)$:
\begin{equation}
{\bf E}\left(\bm{r}\right)={\bf E}^{ext}\left(\bm{r}\right)+i\omega\mu_{0}\int{\bf G}\left(\bm{r}-\bm{r}'\right){\bf J}\left(\bm{r}'\right)d^{3}\bm{r}'.\label{eq:vie}
\end{equation}
The spherical vector wave expansion of ${\bf G}\left(\bm{r}-\bm{r}'\right)$
in terms of (\ref{eq:svwf_M}), (\ref{eq:svwf_N}) is \cite{Tsang2000}:
\begin{align}
{\bf G}\left(\bm{r}-\bm{r}'\right) & =-\frac{1}{k_{b}^{2}}\hat{\bm{e}}_{r}^{T}\hat{\bm{e}}_{r}\delta\left(r-r'\right)\nonumber \\
 & +ik_{b}\begin{cases}
\sum_{nm}\mathcal{{\bf \mathcal{M}}}_{nm}^{1\ast}\left(k_{b}\bm{r}'\right)\mathcal{{\bf \mathcal{M}}}_{nm}^{3}\left(k_{b}\bm{r}\right)+\mathcal{{\bf \mathcal{N}}}_{nm}^{1\ast}\left(k_{b}\bm{r}'\right)\mathcal{{\bf \mathcal{N}}}_{nm}^{3}\left(k_{b}\bm{r}\right), & r'<r,\\
\sum_{nm}\mathcal{{\bf \mathcal{M}}}_{nm}^{3\ast}\left(k_{b}\bm{r}'\right)\mathcal{{\bf \mathcal{M}}}_{nm}^{1}\left(k_{b}\bm{r}\right)+\mathcal{{\bf \mathcal{N}}}_{nm}^{3\ast}\left(k_{b}\bm{r}'\right)\mathcal{{\bf \mathcal{N}}}_{nm}^{1}\left(k_{b}\bm{r}\right), & r'>r.
\end{cases}\label{eq:g0_eig}
\end{align}
where the asterisk $\ast$ stands for complex conjugation. This explicitly
defines operator $\mathfrak{S}_{b}$ in Eq. (\ref{eq:gsm_basis}).
Substitution of the latter expression into the equation (\ref{eq:vie})
shows that the resulting field at any space point is a superposition
of the vector spherical waves together with the singular term existent
in the source region. With a view of simplifying the following derivation
let us introduce the modified field $\tilde{{\bf E}}$, such that
$\tilde{E}_{r}=E_{r}-J_{r}/i\omega\mu_{b}$, $\tilde{E}_{\theta,\varphi}\equiv E_{\theta,\varphi}$.
Therefore, the solution of the volume integral equation can be written
purely as a sum of the regular and outgoing waves
\begin{equation}
\tilde{{\bf E}}\left(\bm{r}\right)={\bf E}^{ext}\left(\bm{r}\right)+\sum_{nm}\left[\tilde{a}_{nm}^{e}\left(r\right){\bf \mathcal{M}}{}_{nm}^{3}\left(k_{b}\bm{r}\right)+\tilde{a}_{nm}^{h}\left(r\right){\bf \mathcal{N}}_{nm}^{3}\left(k_{b}\bm{r}\right)+\tilde{b}_{nm}^{e}\left(r\right){\bf \mathcal{M}}_{nm}^{1}\left(k_{b}\bm{r}\right)+\tilde{b}_{nm}^{h}\left(r\right){\bf \mathcal{N}}_{nm}^{1}\left(k_{b}\bm{r}\right)\right].\label{eq:vol_sol}
\end{equation}
Since the sources of the external field ${\bf E}^{ext}\left(\bm{r}\right)$
lie outside the region of interest where the solution field is to
be evaluated, this external field is also a superposition of vector
spherical waves with constant coefficients being the same as its modified
counterpart, i.e., $\tilde{a}_{nm}^{ext,e,h}\equiv a_{nm}^{ext,e,h}$,
$\tilde{b}_{nm}^{ext,e,h}\equiv b_{nm}^{ext,e,h}$.

Suppose that the region of interest is a spherical layer $R_{1}\leq r\leq R_{2}$.
Then, radially dependent amplitudes in Eq. (\ref{eq:vol_sol}) are
written through weighted spherical harmonics of the source components
\begin{equation}
\tilde{a}_{nm}^{e}\left(r\right)=\tilde{a}_{nm}^{ext,e}\left(r\right)+\intop_{k_{b}R_{1}}^{k_{b}r}\left[\mathcal{J}_{+,nm}\left(x\right)+\mathcal{J}_{-,nm}\left(x\right)\right]j_{n}\left(x\right)x^{2}dx\label{eq:ae_J}
\end{equation}
\begin{equation}
\tilde{b}_{nm}^{e}\left(r\right)=\tilde{b}_{nm}^{ext,e}\left(r\right)+\intop_{k_{b}r}^{k_{b}R_{2}}\left[\mathcal{J}_{+,nm}\left(x\right)+\mathcal{J}_{-,nm}\left(x\right)\right]h_{n}^{(1)}\left(x\right)x^{2}dx\label{eq:be_J}
\end{equation}
\begin{equation}
\tilde{a}_{nm}^{h}\left(r\right)=\tilde{a}_{nm}^{ext,h}\left(r\right)+i\intop_{k_{b}R_{1}}^{k_{b}r}\left\{ \left[\mathcal{J}_{+,nm}\left(x\right)-\mathcal{J}_{-,nm}\left(x\right)\right]\tilde{j}_{n}\left(x\right)+\mathcal{J}_{r,nm}\left(x\right)j_{n}\left(x\right)\right\} xdx\label{eq:ah_J}
\end{equation}
\begin{equation}
\tilde{b}_{nm}^{h}\left(r\right)=\tilde{b}_{nm}^{ext,h}\left(r\right)+i\intop_{k_{b}r}^{k_{b}R_{2}}\left\{ \left[\mathcal{J}_{+,nm}\left(x\right)-\mathcal{J}_{-,nm}\left(x\right)\right]\tilde{h}_{n}^{(1)}\left(x\right)+\mathcal{J}_{r,nm}\left(x\right)h_{n}^{(1)}\left(x\right)\right\} xdx\label{eq:bh_J}
\end{equation}
where
\begin{equation}
\mathcal{J}_{r,nm}\left(x\right)=\frac{\sqrt{n\left(n+1\right)}}{\sqrt{2\pi}}\intop_{0}^{2\pi}\exp\left(-im\varphi\right)d\varphi\intop_{0}^{\pi}\frac{J_{r}\left(x,\theta,\varphi\right)}{-i\omega\varepsilon_{b}}P_{n}^{m}\left(\theta\right)\sin\theta d\theta\label{eq:jr_dec}
\end{equation}
\begin{equation}
\mathcal{J}_{\pm,nm}\left(x\right)=\frac{\sqrt{2n+1}}{2\sqrt{2\pi}}\intop_{0}^{2\pi}\exp\left(-im\varphi\right)d\varphi\intop_{0}^{\pi}\frac{J_{\pm}\left(x,\theta,\varphi\right)}{-i\omega\varepsilon_{b}}d_{m,\pm1}^{n}\left(\theta\right)\sin\theta d\theta\label{eq:jpm_dec}
\end{equation}
Here the explicit expressions for the vector spherical wave functions
(\ref{eq:svwf_M},\ref{eq:svwf_N}) were used.

Within this work Eqs. (\ref{eq:ae_J})-(\ref{eq:bh_J}) serve as a
starting point for derivation of both methods for the complete T-matrix
computation (scattering matrix method) and for T-matrix single column
computation (scattering vector method).

\section{Equations in the homogeneous basis medium}

Following the rationale of the GSM given in Section 2 the basis solution
of the previous section should be supplemented with the generalized
current related to the field as ${\bf J}_{gen}\left(\bm{r}\right)=-i\omega\left[\varepsilon\left(\bm{r}\right)-\varepsilon_{b}\right]{\bf E}\left(\bm{r}\right)$.
Invoking the substitution of the real field with the modified field
one can acquire the following matrix relation
\begin{equation}
{\bf J}_{gen}\left(\bm{r}\right)=-i\omega\left(\begin{array}{ccc}
\Delta\varepsilon\left(\bm{r}\right)/\varepsilon\left(\bm{r}\right) & 0 & 0\\
0 & \Delta\varepsilon\left(\bm{r}\right)/\varepsilon_{b} & 0\\
0 & 0 & \Delta\varepsilon\left(\bm{r}\right)/\varepsilon_{b}
\end{array}\right)\tilde{{\bf E}}\left(\bm{r}\right)\label{eq:src_gen}
\end{equation}
again providing that the vectors are written in the $\left(\hat{\bm{e}}_{r},\hat{\bm{e}}_{+},\hat{\bm{e}}_{-}\right)$
basis and $\Delta\varepsilon\left(\bm{r}\right)=\varepsilon\left(\bm{r}\right)-\varepsilon_{b}$.
Since Eqs. (\ref{eq:jr_dec}), (\ref{eq:jpm_dec}) involve spherical
harmonics of the source, it will be assumed further that the permittivity
functions in Eq. (\ref{eq:src_gen}) admit the spherical harmonic
decomposition
\begin{equation}
\left(\begin{array}{c}
\Delta\varepsilon\left(\bm{r}\right)/\varepsilon\left(\bm{r}\right)\\
\Delta\varepsilon\left(\bm{r}\right)/\varepsilon_{b}
\end{array}\right)=\sum_{nm}\left(\begin{array}{c}
\left[\Delta\varepsilon\left(\bm{r}\right)/\varepsilon\left(\bm{r}\right)\right]_{nm}\\
\left[\Delta\varepsilon\left(\bm{r}\right)/\varepsilon_{b}\right]_{nm}
\end{array}\right)\left(r\right)P_{n}^{m}\left(\theta\right)\exp\left(im\varphi\right)\label{eq:omega_dec}
\end{equation}

Substitution of Eqs. (\ref{eq:src_gen}) and (\ref{eq:omega_dec})
into (\ref{eq:jr_dec}), (\ref{eq:jpm_dec}), and subsequently into
Eqs. (\ref{eq:ae_J})-(\ref{eq:bh_J}) yields the following self-consistent
system of integral equations on the radially dependent amplitudes
of the spherical vector wave decomposition of the unknown modified
field:
\begin{align}
\tilde{a}_{nm}^{e}\left(r\right)=\tilde{a}_{nm}^{ext,e}+ & i\intop_{k_{b}R_{1}}^{k_{b}r}dx\mathcal{V}_{n1}^{(1)}\left(x\right)\sum_{pq}\left[Q_{nm;pq}^{+}\left(x\right)\left(\tilde{a}_{pq}^{e}\mathcal{V}_{p1}^{(3)}\left(x\right)+\tilde{b}_{pq}^{e}\mathcal{V}_{p1}^{(1)}\left(x\right)\right)\right.\nonumber \\
 & \left.+Q_{nm;pq}^{-}\left(x\right)\left(\tilde{a}_{pq}^{h}\mathcal{V}_{p2}^{(3)}\left(x\right)+\tilde{b}_{pq}^{h}\mathcal{V}_{p2}^{(1)}\left(x\right)\right)\right],\label{eq:ae_sol}
\end{align}
\begin{align}
\tilde{b}_{nm}^{e}\left(r\right)=\tilde{b}_{nm}^{ext,e}+ & i\intop_{k_{b}r}^{k_{b}R_{2}}dx\mathcal{V}_{n1}^{(3)}\left(x\right)\sum_{pq}\left[Q_{nm;pq}^{+}\left(\tilde{a}_{pq}^{e}\mathcal{V}_{p1}^{(3)}\left(x\right)+\tilde{b}_{pq}^{e}\mathcal{V}_{p1}^{(1)}\left(x\right)\right)\right.\nonumber \\
 & \left.+Q_{nm;pq}^{-}\left(x\right)\left(\tilde{a}_{pq}^{h}\mathcal{V}_{p2}^{(3)}\left(x\right)+\tilde{b}_{pq}^{h}\mathcal{V}_{p2}^{(1)}\left(x\right)\right)\right],\label{eq:be_sol}
\end{align}
\begin{align}
\tilde{a}_{nm}^{h}\left(r\right) & =\tilde{a}_{nm}^{ext,h}+i\intop_{k_{b}R_{1}}^{k_{b}r}dx\left\{ \mathcal{V}_{n3}^{(1)}\left(x\right)\sum_{pq}Q_{nm;pq}^{r}\left(x\right)\left(\tilde{a}_{pq}^{h}\mathcal{V}_{p3}^{(3)}\left(x\right)+\tilde{b}_{pq}^{h}\mathcal{V}_{p3}^{(1)}\left(x\right)\right)\right.\nonumber \\
 & \left.-\mathcal{V}_{n2}^{(1)}\left(x\right)\sum_{pq}\left[Q_{nm;pq}^{+}\left(x\right)\left(\tilde{a}_{pq}^{h}\mathcal{V}_{p2}^{(3)}\left(x\right)+\tilde{b}_{pq}^{h}\mathcal{V}_{p2}^{(1)}\left(x\right)\right)+Q_{nm;pq}^{-}\left(x\right)\left(\tilde{a}_{pq}^{e}\mathcal{V}_{p1}^{(3)}\left(x\right)+\tilde{b}_{pq}^{e}\mathcal{V}_{p1}^{(1)}\left(x\right)\right)\right]\right\} ,\label{eq:ah_sol}
\end{align}
\begin{align}
\tilde{b}_{nm}^{h}\left(r\right) & =\tilde{b}_{nm}^{ext,h}+i\intop_{k_{b}r}^{k_{b}R_{2}}dx\left\{ \mathcal{V}_{n3}^{(3)}\left(x\right)\sum_{pq}Q_{nm;pq}^{r}\left(x\right)\left(\tilde{a}_{pq}^{h}\mathcal{V}_{p3}^{(3)}\left(x\right)+\tilde{b}_{pq}^{h}\mathcal{V}_{p3}^{(1)}\left(x\right)\right)\right.\nonumber \\
 & \left.-\mathcal{V}_{n2}^{(3)}\left(x\right)\sum_{pq}\left[Q_{nm;pq}^{+}\left(x\right)\left(\tilde{a}_{pq}^{h}\mathcal{V}_{p2}^{(3)}\left(x\right)+\tilde{b}_{pq}^{h}\mathcal{V}_{p2}^{(1)}\left(x\right)\right)+Q_{nm;pq}^{-}\left(x\right)\left(\tilde{a}_{pq}^{e}\mathcal{V}_{p1}^{(3)}\left(x\right)+\tilde{b}_{pq}^{e}\mathcal{V}_{p1}^{(1)}\left(x\right)\right)\right]\right\} .\label{eq:bh_sol}
\end{align}
To attain these relations one has to, first, apply the orthogonality
of the exponential factors, second, utilize the representation of
the integral of three Wigner $d$-functions via the product of two
Clebsch-Gordan coefficients $C_{p,q;n,m}^{u,s}$ \cite{Varshalovich1988},
and, finally, exploit the symmetry relations $C_{p,q;u,s}^{n,m}=\left(-1\right)^{p-q}\sqrt{\left(2n+1\right)/\left(2u+1\right)}C_{p,q;n,-m}^{u,-s}$,
and $C_{p,q;n,m}^{u,s}=\left(-1\right)^{n+p-u}C_{p,-q;n,-m}^{u,-s}$.
These steps yield the following explicit form of the vectors
\begin{equation}
{\bf \mathcal{V}}_{n}^{(1,3)}\left(x\right)=\left(ixz_{n}^{1,3}\left(x\right),\thinspace\tilde{z}_{n}^{1,3}\left(x\right),\thinspace\sqrt{n\left(n+1\right)}z_{n}^{1,3}\left(x\right)\right)^{T},\label{eq:v_def}
\end{equation}
and the matrix elements
\begin{align}
Q_{nm;pq}^{r} & \left(r\right)=\frac{\left(-1\right)^{q}}{\sqrt{2}}\sqrt{2n+1}\sqrt{2p+1}\sum_{u}\left[\frac{\Delta\varepsilon}{\varepsilon}\right]_{u,m-q}\left(r\right)\frac{C_{p,q;n,-m}^{u,m-q}C_{p,0;n,0}^{u,0}}{\sqrt{2u+1}},\nonumber \\
Q_{nm;pq}^{+} & \left(r\right)=\frac{\left(-1\right)^{q}}{\sqrt{2}}\sqrt{2n+1}\sqrt{2p+1}\sum_{n+p-u=even}\left[\frac{\Delta\varepsilon}{\varepsilon_{b}}\right]_{u,m-q}\left(r\right)\frac{C_{p,q;n,-m}^{u,m-q}C_{p,1;n,-1}^{u,0}}{\sqrt{2u+1}},\label{eq:q_def}\\
Q_{nm;pq}^{-} & \left(r\right)=\frac{\left(-1\right)^{q}}{\sqrt{2}}\sqrt{2n+1}\sqrt{2p+1}\sum_{n+p-u=odd}\left[\frac{\Delta\varepsilon}{\varepsilon_{b}}\right]_{u,m-q}\left(r\right)\frac{C_{p,q;n,-m}^{u,m-q}C_{p,1;n,-1}^{u,0}}{\sqrt{2u+1}}.\nonumber 
\end{align}
The summation over the index $u$ in the latter expressions is performed
under the constraint $\max\left(\left|n-p\right|,\left|m-q\right|\right)\leq u\leq n+p$
for nonvanishing Clebsch-Gordan coefficients \cite{Varshalovich1988}.

The equation system (\ref{eq:ae_sol})-(\ref{eq:bh_sol}) is used
further in two ways. First, the next section presents an analysis
of the scattering by a thin inhomogeneous spherical shell, which brings
the core of the mentioned scattering matrix algorithm being an equivalent
of the IIM. Second, this system is solved self-consistently upon discretization
over a finite radial interval as a part of the scattering vector algorithm.
Note that the integrands in (\ref{eq:ae_sol})-(\ref{eq:bh_sol})
do not depend on the radial distance $r$ as opposed to the initial
volume integral equation (\ref{eq:vie}), and $r$ appears only in
the integration limits. This feature will be used below to formulate
a linear summation part of the scattering vector algorithm. Also,
the external field amplitudes do not depend on $r$, as the basis
medium is supposed to be a homogeneous space.

\section{Scattering matrix of a thin spherical shell}

Let us consider the integration region in Eqs. (\ref{eq:ae_sol})-(\ref{eq:bh_sol})
be a thin spherical shell of the thickness $\Delta R=R_{2}-R_{1}\ll R_{1},R_{2}$,
and having the central point $R_{c}=\left(R_{1}+R_{2}\right)/2$.
For the sake of brevity Eqs. (\ref{eq:ae_sol})-(\ref{eq:bh_sol})
can be rewritten in the following matrix-vector form:
\begin{equation}
\begin{array}{c}
\bm{\tilde{a}}_{nm}\left(r\right)=\bm{\tilde{a}}_{nm}^{ext}+\intop_{k_{b}R_{1}}^{k_{b}r}dx\sum_{pq}\left[\bm{F}_{nm,pq}^{aa}\left(x\right)\bm{\tilde{a}}_{pq}\left(x\right)+\bm{F}_{nm,pq}^{ab}\left(x\right)\bm{\tilde{b}}_{pq}\left(x\right)\right]\\
\bm{\tilde{b}}_{nm}\left(r\right)=\bm{\tilde{b}}_{nm}^{ext}+\intop_{k_{b}r}^{k_{b}R_{2}}dx\sum_{pq}\left[\bm{F}_{nm,pq}^{ba}\left(x\right)\bm{\tilde{a}}_{pq}\left(x\right)+\bm{F}_{nm,pq}^{bb}\left(x\right)\bm{\tilde{b}}_{pq}\left(x\right)\right]
\end{array},\thinspace R_{1}\leq r\leq R_{2}\label{eq:ab_sol}
\end{equation}
where $\bm{\tilde{a}}_{nm}=\left(\tilde{a}_{nm}^{e},\thinspace\tilde{a}_{nm}^{h}\right)^{T}$
and $\bm{\tilde{b}}_{nm}=\left(\tilde{b}_{nm}^{e},\thinspace\tilde{b}_{nm}^{h}\right)^{T}$.
The matrix operator $\bm{F}$ can be directly written out explicitly
on the basis of Eqs. (\ref{eq:ae_sol})-(\ref{eq:bh_sol}), though
it is not required for the purpose of this section. Having reliance
on the smallness of $\Delta R$ the integrals can be approximately
evaluated at the shell boundaries using the midpoint rule:
\begin{equation}
\begin{array}{c}
\bm{\tilde{a}}_{nm}\left(k_{b}R_{2}\right)\approx\bm{\tilde{a}}_{nm}^{ext}+ik_{b}\Delta R\sum_{pq}\left[\bm{F}_{nm,pq}^{aa}\left(k_{b}R_{c}\right)\bm{\tilde{a}}_{pq}\left(k_{b}R_{c}\right)+\bm{F}_{nm,pq}^{ab}\left(k_{b}R_{c}\right)\bm{\tilde{b}}_{pq}\left(k_{b}R_{c}\right)\right],\\
\bm{\tilde{b}}_{nm}\left(k_{b}R_{1}\right)\approx\bm{\tilde{b}}_{nm}^{ext}+ik_{b}\Delta R\sum_{pq}\left[\bm{F}_{nm,pq}^{ba}\left(k_{b}R_{c}\right)\bm{\tilde{a}}_{pq}\left(k_{b}R_{c}\right)+\bm{F}_{nm,pq}^{bb}\left(k_{b}R_{c}\right)\bm{\tilde{b}}_{pq}\left(k_{b}R_{c}\right)\right],
\end{array}\label{eq:approx_sol_bound}
\end{equation}
owing the accuracy of $O\left(\left(\Delta R\right)^{3}\right)$.
Amplitude vectors evaluated at the central point $\bm{\tilde{a}}_{pq}\left(k_{b}R_{c}\right)$,
$\bm{\tilde{b}}_{pq}\left(k_{b}R_{c}\right)$, which appear in the
right-hand sides of the latter Eqs. (\ref{eq:approx_sol_bound}),
can be expressed through another integration when the left-hand side
of Eq. (\ref{eq:ab_sol}) is evaluated at $r=R_{c}$ by aids of the
rectangle rule:
\begin{equation}
\begin{array}{c}
\bm{\tilde{a}}_{nm}\left(k_{b}R_{c}\right)=\bm{\tilde{a}}_{nm}^{ext}+\frac{1}{2}ik_{b}\Delta R\sum_{pq}\left[\bm{F}_{nm,pq}^{aa}\left(k_{b}R_{c}\right)\bm{\tilde{a}}_{pq}\left(k_{b}R_{c}\right)+\bm{F}_{nm,pq}^{ab}\left(k_{b}R_{c}\right)\bm{\tilde{b}}_{pq}\left(k_{b}R_{c}\right)\right]+O\left(\left(\Delta R\right)^{2}\right)\\
\bm{\tilde{b}}_{nm}\left(k_{b}R_{c}\right)=\bm{\tilde{b}}_{nm}^{ext}+\frac{1}{2}ik_{b}\Delta R\sum_{pq}\left[\bm{F}_{nm,pq}^{ba}\left(k_{b}R_{c}\right)\bm{\tilde{a}}_{pq}\left(k_{b}R_{c}\right)+\bm{F}_{nm,pq}^{bb}\left(k_{b}R_{c}\right)\bm{\tilde{b}}_{pq}\left(k_{b}R_{c}\right)\right]+O\left(\left(\Delta R\right)^{2}\right)
\end{array}\label{eq:approx_sol_center}
\end{equation}
This self-consistent system being solved via matrix inversion by neglecting
$O\left(\left(\Delta R\right)^{2}\right)$ terms, the solution should
be substituted into Eq. (\ref{eq:approx_sol_bound}) to yield the
unknown amplitudes at the shell boundaries. However, the inversion
would give an excessive accuracy relative to $\Delta R$ powers, and
therefore one can directly substitute the unknown amplitudes in the
right-hand parts of Eqs. (\ref{eq:approx_sol_bound}) with the external
field amplitudes to keep $O\left(\left(\Delta R\right)^{2}\right)$
accuracy. Formally, these amplitudes of the external field can be
written as if they had been also evaluated at the shell boundaries
as they do not depend on $r$:
\begin{equation}
\begin{array}{c}
\bm{\tilde{a}}_{nm}\left(k_{b}R_{2}\right)=\sum_{pq}\left\{ \left[\delta_{np}\delta_{mq}+ik_{b}\Delta R\bm{F}_{nm,pq}^{aa}\left(k_{b}R_{c}\right)\right]\bm{\tilde{a}}_{pq}^{ext}\left(k_{b}R_{1}\right)+ik_{b}\Delta R\bm{F}_{nm,pq}^{ab}\left(k_{b}R_{c}\right)\bm{\tilde{b}}_{pq}^{ext}\left(k_{b}R_{2}\right)\right\} \\
\bm{\tilde{b}}_{nm}\left(k_{b}R_{1}\right)=\sum_{pq}\left\{ ik_{b}\Delta R\bm{F}_{nm,pq}^{ba}\left(k_{b}R_{c}\right)\bm{\tilde{a}}_{pq}^{ext}\left(k_{b}R_{1}\right)+\left[\delta_{np}\delta_{mq}+ik_{b}\Delta R\bm{F}_{nm,pq}^{bb}\left(k_{b}R_{c}\right)\right]\bm{\tilde{b}}_{pq}^{ext}\left(k_{b}R_{2}\right)\right\} 
\end{array}\label{eq:approx_sol}
\end{equation}
The named excessive inversion is not omitted in the method of \cite{Johnson1988},
though the possibility of formulating a procedure with only one matrix
inversion per step is noted in \cite{Doicu2018}.

An operator transforming the external field amplitudes to the scattered
field for the considered spherical shell amplitudes can be rewritten
in the matrix form:
\begin{equation}
\left(\begin{array}{c}
\bm{\tilde{a}}_{nm}\left(R+\Delta R/2\right)\\
\bm{\tilde{b}}_{nm}\left(R-\Delta R/2\right)
\end{array}\right)=\sum_{pq}\left(\begin{array}{cc}
S_{nm,pq}^{11}\left(R,\thinspace\Delta R\right) & S_{nm,pq}^{12}\left(R,\thinspace\Delta R\right)\\
S_{nm,pq}^{21}\left(R,\thinspace\Delta R\right) & S_{nm,pq}^{22}\left(R,\thinspace\Delta R\right)
\end{array}\right)\left(\begin{array}{c}
\bm{\tilde{b}}_{nm}^{ext}\left(R+\Delta R/2\right)\\
\bm{\tilde{a}}_{nm}^{ext}\left(R-\Delta R/2\right)
\end{array}\right)\label{eq:S-mat}
\end{equation}
We will refer to this matrix as the scattering matrix in an analogy
with the planar geometry case \cite{Shcherbakov2018}.

Inspection of Eq. (\ref{eq:S-mat}) reveals that the components $S_{nm,pq}^{11}$,
$S_{nm,pq}^{22}$ can be viewed as generalized reflection coefficients,
and the components $S_{nm,pq}^{12}$, $S_{nm,pq}^{21}$ -- as generalized
transmission coefficients. The Waterman T-matrix corresponds to the
block $S^{11}=ik_{b}\Delta R\bm{F}^{ab}\left(k_{b}R_{c}\right)$.
Explicit expressions for the S-matrix components follow directly from
Eqs. (\ref{eq:ae_sol})-(\ref{eq:bh_sol}), and are listed in the
Appendix A.

\section{Scattering matrix algorithm}

The approximate scattering matrix of a spherical shell derived above
depends both on the radius and the thickness of this shell. It will
be denoted as $S\left(R,\Delta R\right)$ to distinguish it from the
T-matrix $S^{11}\left(R\right)$ of a scattering medium bounded by
the sphere $r=R$. To use the result of the previous section one has
to provide a calculation algorithm for $S^{11}\left(R+\Delta R\right)$
given the matrix $S\left(R+\Delta R/2,\Delta R\right)$. Such algorithm
was formulated in \cite{Johnson1988} using the invariant imbedding
method. It is derived below by means of some intuitive observations
analogous to the superposition T-matrix method given by \cite{Wriedt2018,Doicu2018}.

Let us consider a spherical scattering region of radius $R$ with
known generalized reflection coefficient matrix $S^{11}\left(R\right)$,
or the T-matrix, as Fig. 2 demonstrates. When illuminated by a wave,
which coefficients in the decomposition into the regular vector spherical
wave functions are $\bm{\tilde{b}}^{ext}$, the coefficient vector
of the outgoing scattered spherical waves is found by the matrix-vector
multiplication $\bm{\tilde{a}}^{sca}=S^{11}\left(R\right)\bm{\tilde{b}}^{ext}$.
Then, let the region be surrounded by a spherical shell of thickness
$\Delta R\ll R$ with known approximate scattering matrix $S\left(R_{s},\thinspace\Delta R\right)$,
$R_{s}=R+\Delta R/2$ derived above. Denote self-consistent amplitudes
of the electromagnetic field at radius $r=R$ as $\bm{\tilde{a}}\left(R\right)$,
and $\bm{\tilde{b}}\left(R\right)$. By definition of the scattering
matrix these amplitudes should meed the following relations:
\begin{equation}
\begin{array}{c}
\bm{\tilde{a}}\left(R\right)=S^{11}\left(R\right)\bm{\tilde{b}}\left(R\right)\\
\bm{\tilde{b}}\left(R\right)=S^{22}\left(R_{s},\thinspace\Delta R\right)\bm{\tilde{a}}\left(R\right)+S^{12}\left(R_{s},\thinspace\Delta R\right)\bm{\tilde{b}}^{ext}
\end{array}\label{eq:shell_eq_sys}
\end{equation}
This yields the explicit expression for the unknown amplitude vector:
\begin{equation}
\bm{\tilde{a}}\left(R\right)=S^{11}\left(R\right)\left[I-S^{22}\left(R_{s},\thinspace\Delta R\right)S^{11}\left(R\right)\right]^{-1}S^{12}\left(R_{s},\thinspace\Delta R\right)\bm{\tilde{b}}^{ext}\label{eq:shell_ampl}
\end{equation}
where $I$ is the identity matrix. The amplitude vector of the scattered
field from the one hand is $\bm{\tilde{a}}^{sca}=S^{11}\left(R_{s},\thinspace\Delta R\right)\bm{\tilde{b}}^{ext}+S^{12}\left(R_{s},\thinspace\Delta R\right)\bm{\tilde{a}}\left(R\right)$,
and from the other hand, it should be$\bm{\tilde{a}}^{sca}=S^{11}\left(R+\Delta R\right)\bm{\tilde{b}}^{ext}$.
Thus, the desired reflection matrix of the compounded scatterer is
\begin{equation}
S^{11}\left(R+\Delta R\right)=S^{11}\left(R_{s},\thinspace\Delta R\right)+S^{12}\left(R_{s},\thinspace\Delta R\right)S^{11}\left(R\right)\left[I-S^{22}\left(R_{s},\thinspace\Delta R\right)S^{11}\left(R\right)\right]^{-1}S^{12}\left(R_{s},\thinspace\Delta R\right)\label{eq:s_matrix _alg}
\end{equation}

\begin{figure}
\begin{centering}
\includegraphics[scale=0.6]{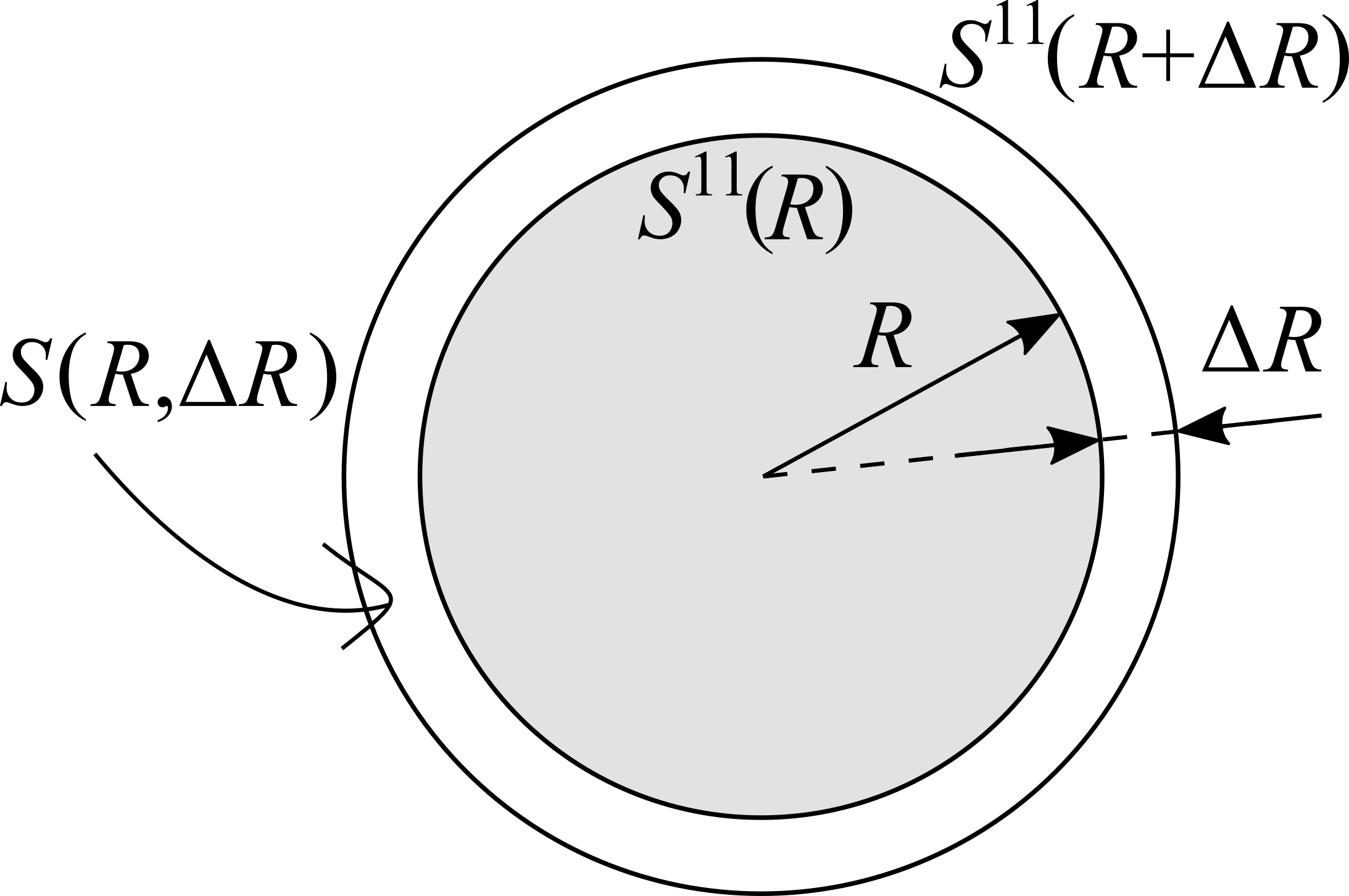}
\par\end{centering}
\caption{To the derivation of the scattering matrix algorithm.}

\end{figure}

A full calculation algorithm implementing Eq. (\ref{eq:s_matrix _alg})
is also similar to what is described in \cite{Johnson1988}. Owing
a homogeneous scattering particle with a continuous closed surface
$\partial\Omega_{p}$ one should, first, choose a center of a spherical
coordinate system, second, choose an inner and an outer spherical
interfaces of radii $R_{in}$, $R_{out}$ centered at the origin,
the first one being inscribed inside the particle, and the second
one being circumscribed around it. The particle surface appears to
be enclosed in the spherical layer. This partitioning is schematically
shown in Fig. 3a,b. The permittivity inside the inscribed sphere is
a constant being equal to the particle permittivity $\varepsilon_{p}$.
Similarly, the permittivity of the region $r>R_{out}$ is the surrounding
medium permittivity $\varepsilon_{m}$. The basis permittivity $\varepsilon_{b}$
to be ascribed to the region $R_{in}\le r\le R_{out}$ is essentially
a free parameter of the method and is a matter of choice. Ideally
it should not affect any physical quantities at the output of the
method, and its influence on simulations and results will be discussed
below.

A starting point of the algorithm is the diagonal matrix $S^{11}\left(R_{in}\right)$
filled with the Mie scattering coefficients obtained from the continuity
boundary conditions at the interface $r=R_{in}$ separating media
with permittivities $\varepsilon_{p}$ and $\varepsilon_{b}$ \cite{Bohren2007}.
Explicitly,
\begin{equation}
\begin{array}{c}
S_{nm}^{11,ee}\left(R_{in}\right)=\dfrac{j_{n}\left(k_{p}R_{in}\right)\tilde{j}_{n}\left(k_{b}R_{in}\right)-j_{n}\left(k_{b}R_{in}\right)\tilde{j}_{n}\left(k_{p}R_{in}\right)}{h_{n}^{(1)}\left(k_{b}R_{in}\right)\tilde{j}_{n}\left(k_{p}R_{in}\right)-j_{n}\left(k_{p}R_{in}\right)\tilde{h}_{n}^{(1)}\left(k_{b}R_{in}\right)}\\
S_{nm}^{11,hh}\left(R_{in}\right)=\dfrac{j_{n}\left(k_{p}R_{in}\right)\tilde{j}_{n}\left(k_{b}R_{in}\right)-\dfrac{\varepsilon_{b}}{\varepsilon_{p}}j_{n}\left(k_{b}R_{in}\right)\tilde{j}_{n}\left(k_{p}R_{in}\right)}{\dfrac{\varepsilon_{b}}{\varepsilon_{p}}h_{n}^{(1)}\left(k_{b}R_{in}\right)\tilde{j}_{n}\left(k_{p}R_{in}\right)-j_{n}\left(k_{p}R_{in}\right)\tilde{h}_{n}^{(1)}\left(k_{b}R_{in}\right)}
\end{array}\label{eq:mie_smatrix}
\end{equation}
and $S_{nm}^{11,eh}\left(R_{in}\right)=S_{nm}^{11,he}\left(R_{in}\right)=0$.
Upper indices ``$e$'' and ``$h$'' correspond to the polarization
of input and output waves, and $k_{p,b}=\omega\sqrt{\varepsilon_{p,b}\mu_{0}}$.
Then, the layer $R_{in}<r<R_{out}$ of permittivity $\varepsilon_{b}$
should be divided into a number of thin shells, while the scattering
matrix of each shell is explicitly given in Appendix A.

After $S^{11}$ matrix is accumulated by means of Eq. (\ref{eq:s_matrix _alg})
for the increasing radial distance, a final multiplication by the
scattering matrix of the spherical interface $r=R_{out}$ separating
the basis and the outer medium should be made. To give this matrix
explicitly we, first, fix the field decomposition in the vicinity
of this interface. For $r=R_{out}-0$ this decomposition is the one
used in all above derivations -- into a superposition of regular
and outgoing spherical waves (see Eq. (\ref{eq:vol_sol})). For $r>R_{out}$
it is convenient to have the fields decomposed into a set of incoming
and outgoing waves 
\begin{equation}
{\bf E}\left(\bm{r}\right)=\sum_{nm}\left[a_{nm}^{e}{\bf \mathcal{M}}{}_{nm}^{3}\left(k_{s}\bm{r}\right)+a_{nm}^{h}{\bf \mathcal{N}}_{nm}^{3}\left(k_{s}\bm{r}\right)+c_{nm}^{e}{\bf \mathcal{M}}_{nm}^{2}\left(k_{s}\bm{r}\right)+c_{nm}^{h}{\bf \mathcal{N}}_{nm}^{2}\left(k_{s}\bm{r}\right)\right],\thinspace r>R_{out}.\label{eq:in-out-decomposition}
\end{equation}
in order to define the incoming field via coefficients $c_{nm}^{e,h}$.
Here $k_{s}=\omega\sqrt{\varepsilon_{s}\mu_{0}}$ is the wavenumber
in the surrounding non-absorbing medium, and the upper index $"2"$
indicates that the corresponding spherical wave functions are written
via the spherical Henkel functions of the second kind $z_{n}^{2}\equiv h_{n}^{(2)}$.
Analogously to the Mie scattering scenario the boundary conditions
yield a corresponding diagonal scattering matrix components. These
components are specified in Appendix B.

\section{Field solution in a basis spherical layer}

Now let us return to the basis Eqs. (\ref{eq:ae_sol})-(\ref{eq:bh_sol})
and use them for the self-consistent calculation of T-matrix single
columns. In other words, for calculation of a response amplitude vector
to a given excitation field. The basis equations now should be extended
to yield a solution in the basis spherical layer $R_{in}\leq r\leq R_{out}$
taking into account multiple reflections at the layer interfaces (see
Fig. 3b). Instead of constructing a corresponding Green's tensor we
will directly construct a field solution on the basis of the derived
solution for the homogeneous space.

Let the reader recall the definition of the inner and the outer spherical
interfaces bounding the scattering particle surface $\partial\Omega_{p}$,
Fig. 3a, and put $R_{1}=R_{in}$, $R_{2}=R_{out}$ in (\ref{eq:ae_sol})-(\ref{eq:bh_sol}).
Then, the self-consistent field can be searched in form (e.g., see
\cite{Li1994})
\begin{equation}
\begin{array}{c}
\tilde{{\bf E}}\left(\bm{r}\right)=\sum_{nm}\left[C_{nm}^{ae}\tilde{a}_{nm}^{e}\left(R_{out}\right)+C_{nm}^{be}\tilde{b}_{nm}^{e}\left(R_{in}\right)\right]\mathcal{M}_{nm}^{1}\left(k_{p}\bm{r}\right)\\
+\left[C_{nm}^{ah}\tilde{a}_{nm}^{h}\left(R_{out}\right)+C_{nm}^{bh}\tilde{b}_{nm}^{h}\left(R_{in}\right)\right]\mathcal{N}_{nm}^{1}\left(k_{p}\bm{r}\right)
\end{array}\label{eq:basis_sol_layer_in}
\end{equation}
for $r<R_{in}$,
\begin{equation}
\begin{array}{c}
\tilde{{\bf E}}\left(\bm{r}\right)=\sum_{nm}\left[\tilde{a}_{nm}^{e}\left(r\right)+A_{nm}^{ae}\tilde{a}_{nm}^{e}\left(R_{out}\right)+A_{nm}^{be}\tilde{b}_{nm}^{e}\left(R_{in}\right)\right]\mathcal{M}_{nm}^{3}\left(k_{b}\bm{r}\right)\\
+\left[\tilde{a}_{nm}^{h}\left(r\right)+A_{nm}^{ah}\tilde{a}_{nm}^{h}\left(R_{out}\right)+A_{nm}^{bh}\tilde{b}_{nm}^{h}\left(R_{in}\right)\right]\mathcal{N}_{nm}^{3}\left(k_{b}\bm{r}\right)\\
+\left[\tilde{b}_{nm}^{e}\left(r\right)+B_{nm}^{ae}\tilde{a}_{nm}^{e}\left(R_{out}\right)+B_{nm}^{be}\tilde{b}_{nm}^{e}\left(R_{in}\right)\right]\mathcal{M}_{nm}^{1}\left(k_{b}\bm{r}\right)\\
+\left[\tilde{b}_{nm}^{h}\left(r\right)+B_{nm}^{ah}\tilde{a}_{nm}^{h}\left(R_{out}\right)+B_{nm}^{bh}\tilde{b}_{nm}^{h}\left(R_{in}\right)\right]\mathcal{N}_{nm}^{1}\left(k_{b}\bm{r}\right)
\end{array}\label{eq:basis_sol_layer_ins}
\end{equation}
 for \textbf{$R_{in}\leq r\leq R_{out}$}, and
\begin{equation}
\begin{array}{c}
\tilde{{\bf E}}\left(\bm{r}\right)=\sum_{nm}\left[D_{nm}^{ae}\tilde{a}_{nm}^{e}\left(R_{out}\right)+D_{nm}^{be}\tilde{b}_{nm}^{e}\left(R_{in}\right)\right]\mathcal{M}_{nm}^{3}\left(k_{m}\bm{r}\right)\\
+\left[D_{nm}^{ah}\tilde{a}_{nm}^{e}\left(R_{out}\right)+D_{nm}^{bh}\tilde{b}_{nm}^{e}\left(R_{in}\right)\right]\mathcal{N}_{nm}^{3}\left(k_{m}\bm{r}\right)
\end{array}\label{eq:basis_sol_layer_out}
\end{equation}
for $r>R_{out}$. Similar expressions for the magnetic field follow
directly from the first of Maxwell's Eq. (\ref{eq:maxwell}) and transformation
relations (\ref{eq:MN_NM}). Note, that according to (\ref{eq:ae_sol})-(\ref{eq:bh_sol})
$\tilde{a}_{nm}^{e,h}\left(R_{in}\right)=\tilde{b}_{nm}^{e,h}\left(R_{out}\right)=0$.
The external field here is no more the field of the homogeneous space,
but rather the self-consistent field of the spherical particle of
radius $r=R_{in}$ and permittivity $\varepsilon_{p}$ covered by
the spherical layer $R_{in}\leq r\leq R_{out}$ of permittivity $\varepsilon_{b}$,
and placed in the medium of permittivity $\varepsilon_{s}$.

Unknown sets of coefficients $A$, $B$, $C$, and $D$ are found
from the continuity of the tangential field components at the interfaces
$r=R_{in},R_{out}$ analogously to the Mie theory and derivations
of Appendix B. They can be expressed via the S-matrices of the inner
and the outer interfaces introduced within the scattering matrix algorithm
(see Eq. (\ref{eq:mie_smatrix}) and Appendix B). The resulting formulas
explicitly write
\begin{equation}
A_{nm}^{a\star}\tilde{a}_{nm}^{\star}\left(R_{out}\right)+A_{nm}^{b\star}\tilde{b}_{nm}^{\star}\left(R_{in}\right)=S_{nm}^{11,\star\star}\left(R_{in}\right)\frac{S_{nm}^{22,\star\star}\left(R_{out}\right)\tilde{a}_{nm}^{\star}\left(R_{out}\right)+\tilde{b}_{nm}^{\star}\left(R_{in}\right)}{1-S_{nm}^{11,\star\star}\left(R_{in}\right)S_{nm}^{22,\star\star}\left(R_{out}\right)}\label{eq:coef_A}
\end{equation}
\begin{equation}
B_{nm}^{a\star}\tilde{a}_{nm}^{\star}\left(R_{out}\right)+B_{nm}^{b\star}\tilde{b}_{nm}^{\star}\left(R_{in}\right)=S_{nm}^{22,\star\star}\left(R_{out}\right)\frac{\tilde{a}_{nm}^{\star}\left(R_{out}\right)+S_{nm}^{11,\star\star}\left(R_{in}\right)\tilde{b}_{nm}^{\star}\left(R_{in}\right)}{1-S_{nm}^{11,\star\star}\left(R_{in}\right)S_{nm}^{22,\star\star}\left(R_{out}\right)}\label{eq:coef_B}
\end{equation}
\begin{equation}
C_{nm}^{a\ast}\tilde{a}_{nm}^{\star}\left(R_{out}\right)+C_{nm}^{b\star}\tilde{b}_{nm}^{\star}\left(R_{in}\right)=S_{nm}^{21,\star\star}\left(R_{in}\right)\frac{S_{nm}^{22,\star\star}\left(R_{out}\right)\tilde{a}_{nm}^{\star}\left(R_{out}\right)+\tilde{b}_{nm}^{\star}\left(R_{in}\right)}{1-S_{nm}^{11,\star\star}\left(R_{in}\right)S_{nm}^{22,\star\star}\left(R_{out}\right)}\label{eq:coef_C}
\end{equation}
\begin{equation}
D_{nm}^{a\star}\tilde{a}_{nm}^{\star}\left(R_{out}\right)+D_{nm}^{b\star}\tilde{b}_{nm}^{\star}\left(R_{in}\right)=S_{nm}^{12,\star\star}\left(R_{out}\right)\frac{S_{nm}^{22,\star\star}\left(R_{out}\right)\tilde{a}_{nm}^{\star}\left(R_{out}\right)+\tilde{b}_{nm}^{\star}\left(R_{in}\right)}{1-S_{nm}^{11,\star\star}\left(R_{in}\right)S_{nm}^{22,\star\star}\left(R_{out}\right)}\label{eq:coef_D}
\end{equation}
where the star $\star$ can be either ``$e$'', or ``$h$''. Also
used here the transmission coefficients of the inner interface are
\begin{equation}
\begin{array}{c}
S_{nm}^{21,ee}\left(R_{in}\right)=\dfrac{i}{k_{b}R_{in}}\dfrac{1}{j_{n}\left(k_{p}R_{in}\right)\tilde{h}_{n}^{(1)}\left(k_{b}R_{in}\right)-h_{n}^{(1)}\left(k_{b}R_{in}\right)\tilde{j}_{n}\left(k_{p}R_{in}\right)}\\
S_{nm}^{21,hh}\left(R_{in}\right)=\dfrac{i}{k_{p}R_{in}}\dfrac{1}{j_{n}\left(k_{p}R_{in}\right)\tilde{h}_{n}^{(1)}\left(k_{b}R_{in}\right)-\dfrac{\varepsilon_{b}}{\varepsilon_{p}}h_{n}^{(1)}\left(k_{b}R_{in}\right)\tilde{j}_{n}\left(k_{p}R_{in}\right)}
\end{array}\label{eq:S_inner_transm}
\end{equation}
Eqs. (\ref{eq:basis_sol_layer_in})-(\ref{eq:basis_sol_layer_out})
with coefficients (\ref{eq:coef_A})-(\ref{eq:coef_D}) and radially
dependent amplitudes $\tilde{a}_{nm}^{e,h}\left(r\right)$, $\tilde{b}_{nm}^{e,h}\left(r\right)$
defined by Eqs. (\ref{eq:ae_sol})-(\ref{eq:bh_sol}) are the solution
sought in the case of the spherical layer basis medium.

\begin{figure}
\begin{centering}
\includegraphics[scale=0.6]{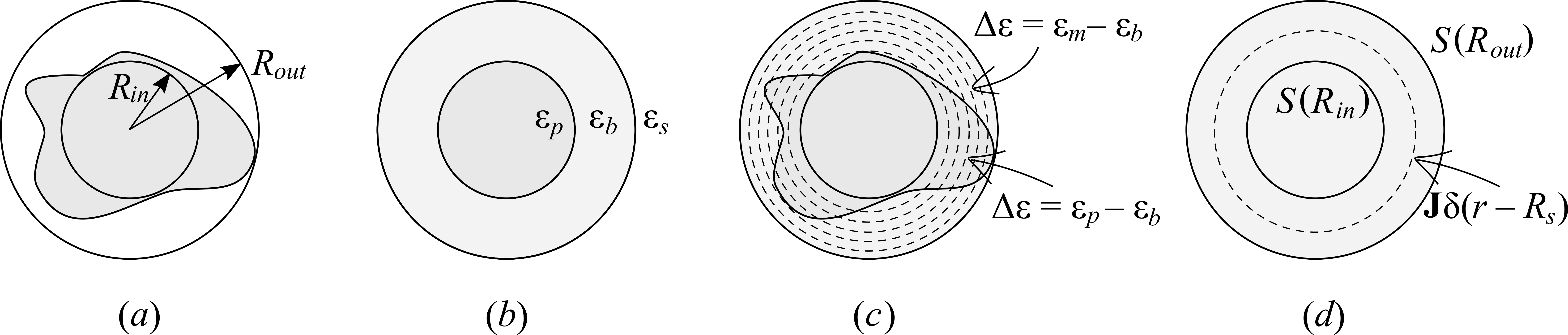}
\par\end{centering}
\caption{Illustrations to the scattering matrix and the scattering vector algorithms.
a) inscribed and ascribed spherical surfaces; b) basis medium for
the scattering vector algorithm consisting of a core of a permittivity
equal to the particle permittivity, a spherical layer of a permittivity
being a free parameter of the methods, and a surrounding medium; c)
slicing of the basis layer into a set of thin spherical shells; d)
a radial ``delta''-source located inside the basis layer.}
\end{figure}

It might be interesting to note that Eqs. (\ref{eq:coef_A})-(\ref{eq:coef_D})
can be derived with the aids of considerations similar to those given
while developing the scattering matrix algorithm in the previous section.
Consider a spherical layer (Fig. ...) $R_{in}\leq r\leq R_{out}$
together with the volume ``delta''-source inside ${\bf J}=\delta(r-R_{s}){\bf J}\left(\theta,\varphi\right)$,
$R_{in}<R_{s}<R_{out}$. Let the field emitted by the source be decomposed
into the spherical vector waves having amplitude vectors $\bm{b_{J}}$
for $r<R_{s}$ (being regular wave functions), and $\bm{a_{J}}$ for
$r>R_{s}$ (being outgoing wave functions). These are amplitudes which
the source would emit in the absence of the interfaces $r=R_{in,out}$.
To find self-consistent amplitudes in the presence of the interfaces
let us use the S-matrix relations at both interfaces $S\left(R_{in,out}\right)$:
\begin{equation}
\begin{array}{c}
\bm{b}=\bm{b_{J}}+S^{11}\left(R_{out}\right)\bm{a}\\
\bm{a}=\bm{a_{J}}+S^{22}\left(R_{in}\right)\bm{b}
\end{array}\label{eq:sys-source-in-layer}
\end{equation}
where $\bm{a}$ and $\bm{b}$ are unknown amplitude vectors of the
self-consistent filed inside the layer. Solution of these equations
is
\begin{equation}
\begin{array}{c}
\bm{b}=\bm{b_{J}}+S^{11}\left(R_{out}\right)\left[I-S^{22}\left(R_{in}\right)S^{11}\left(R_{out}\right)\right]^{-1}\left[\bm{a_{J}}+S^{22}\left(R_{in}\right)\bm{b_{J}}\right]\\
\bm{a}=\bm{a_{J}}+S^{22}\left(R_{in}\right)\left[I-S^{11}\left(R_{out}\right)S^{22}\left(R_{in}\right)\right]^{-1}\left[\bm{b_{J}}+S^{11}\left(R_{out}\right)\bm{a_{J}}\right]
\end{array}\label{eq:sol-source-in-layer}
\end{equation}
Bearing in mind the fact that scattering matrices of spherical interfaces
are diagonal, Eqs. (\ref{eq:sol-source-in-layer}) immediately become
(\ref{eq:coef_A}), (\ref{eq:coef_B}). Amplitude vectors inside the
sphere $r<R_{1}$, and in the outer space $r>R_{2}$ are found via
S-matrix relations on the basis of this self-consistent field as $\bm{b}|_{r<R_{1}}=S^{21}\left(R_{in}\right)\bm{b}$,
$\bm{a}|_{r>R_{2}}=S^{12}\left(R_{out}\right)\bm{b}$.

\section{Scattering vector algorithm}

To solve the equations derived in the previous section the spherical
layer should be divided into $N_{s}$ thin spherical shells of thickness
$\Delta R=\left(R_{out}-R_{in}\right)/N_{s}$ (Fig. 3c) to approximate
the integration by finite sums (here the mid-point rule is applied).
Upon truncation of infinite series of the spherical wave functions
one runs into a finite self-consistent linear equation system on unknown
field amplitudes. This system can be written as follows:
\begin{equation}
\bm{V}=\left({\bf I}-\mathcal{P}\mathcal{R}\right)^{-1}\bm{V}^{inc}\label{eq:svector_solution}
\end{equation}
where vectors $\bm{V}$, $\bm{V}^{inc}$ contain unknown and incident
amplitudes in all shells $\tilde{a}_{nm,k}^{e,h}=\tilde{a}_{nm}^{e,h}\left(r_{k}\right)$,
$\tilde{b}_{nm,k}^{e,h}=\tilde{b}_{nm}^{e,h}\left(r_{k}\right)$,
$r_{k}=R_{in}+\left(k+1/2\right)\Delta R$, $k=0,\thinspace1,\thinspace\dots,\thinspace N_{s}-1$.
Matrix operator $\mathcal{R}$ describes the scattering in each thin
shell. It is diagonal relative to the index $k$ enumerating shells,
and explicitly is given by Eq. (\ref{eq:S-mat}) with S-matrix components
listed in Appendix A. The second operator $\mathcal{P}$ corresponds
to propagation of the vector spherical waves between different shells,
and implies the weighted summation of the output of the operator $\mathcal{R}$
in accordance with Eqs. (\ref{eq:basis_sol_layer_in})-(\ref{eq:basis_sol_layer_out}).
This summation can be organized with two loops, so that the resulting
algorithm can be formulated as follows:
\begin{enumerate}
\item Choose the truncation number for spherical harmonics $N=\max n$,
and pre-calculate coefficient matrices $A$, $B$, $C$, and $D$
on the basis of Eqs. (\ref{eq:coef_A})-(\ref{eq:coef_D}).
\item Choose a basis layer, its subdivision into shells, and calculate spherical
harmonic transformation of the permittivity, Eq. (\ref{eq:omega_dec}),
in each shell. Pre-calculate S-matrix components for all shells on
the basis of Eqs. (\ref{eq:v_def}), (\ref{eq:q_def}), and Appendix
A.
\item Solve Eq. (\ref{eq:svector_solution}) by means of an iterative method
like the Bi-conjugate Gradient or the Generalized Minimal Residual
method. At each iteration step do the following:
\begin{enumerate}
\item multiply amplitudes in each shell by the corresponding scattering
matrices in accordance with Eq. (\ref{eq:S-mat}) and Appendix A.
\item loop over shells and store the partial sums for each shell: $\tilde{\bm{a}}_{nm,k}^{sum}=\sum_{j=0}^{k}\tilde{\bm{a}}_{nm,j}$,
$\tilde{\bm{b}}_{nm,k}^{sum}=\sum_{j=k}^{N_{s}-1}\tilde{\bm{b}}_{nm,j}$.
Then the accumulated amplitudes $\tilde{\bm{a}}_{nm,N_{s}-1}^{sum}$
and $\tilde{\bm{b}}_{nm,0}^{sum}$ should be multiplied by the weight
factors according to Eqs. (\ref{eq:coef_A})-(\ref{eq:coef_D}) and
added to the stored amplitudes in each shell as Eq. (\ref{eq:basis_sol_layer_ins})
requires.
\end{enumerate}
\end{enumerate}

\section{Implementation details}

The described methods were implemented in C++ and compiled under Windows
using the MS Visual Studio compiler. In particular, spherical Bessel
functions are calculated on the basis of \cite{Babushkina1988}, associated
Legendre polynomials -- on the basis of \cite{Gil2007}, Clebsch-Gordan
coefficients -- on the basis of \cite{Mishchenko1991}. An important
feature of the referenced algorithms is that they allow calculating
whole sets of required functions for one of their indices at a single
iteration cycle, which substantially saves computation time. Both
the scattering matrix and the scattering vector algorithms require
evaluation of factors in Eqs. (\ref{eq:q_def}), which do not depend
on a particular permittivity function and on radial distance. Namely,
matrices
\begin{equation}
\begin{array}{c}
\hat{Q}_{nm;pq,u}^{r}=\dfrac{\left(-1\right)^{q}}{\sqrt{2}}\sqrt{2n+1}\sqrt{2p+1}\dfrac{C_{p,q;n,-m}^{u,m-q}C_{p,0;n,0}^{u,0}}{\sqrt{2u+1}}\\
\hat{Q}_{nm;pq,u}^{\pm}=\dfrac{\left(-1\right)^{q}}{\sqrt{2}}\sqrt{2n+1}\sqrt{2p+1}\dfrac{C_{p,q;n,-m}^{u,m-q}C_{p,1;n,-1}^{u,0}}{\sqrt{2u+1}}
\end{array}\label{eq:q_matr}
\end{equation}
are pre-calculated, stored, and then used for finding the scattering
matrix elements in each shell. In case of large-scale computations
it seems to be reasonable to create a library of matrix elements (\ref{eq:q_matr})
for them to be readily available. Within the scattering vector algorithm
vectors (\ref{eq:v_def}), transmission and reflection coefficients
for the inner and the outer spherical interfaces are also pre-calculated.

The following examples concern scattering by spheroids, and general
polyhedral particles. The spheroidal shape particles with the axis
of revolution coinciding with the coordinate axis $Z$ allow for analytic
formulas for the spherical harmonic decomposition of the dielectric
function:
\begin{align}
\left[\mathcal{E}\right]_{nm}\left(r\right) & =\frac{1}{2\pi}\intop_{0}^{\pi}\mathcal{E}\left(r,\theta\right)P_{n}^{m}\left(\theta\right)\sin\theta d\theta\intop_{0}^{2\pi}\exp\left(-im\varphi\right)d\varphi\nonumber \\
 & =\delta_{m0}\mathcal{E}_{1}\left[\intop_{0}^{\Theta\left(r\right)}P_{n}\left(\theta\right)\sin\theta d\theta+\intop_{\pi-\Theta\left(r\right)}^{\pi}P_{n}\left(\theta\right)\sin\theta d\theta\right]+\delta_{m0}\mathcal{E}_{2}\intop_{\Theta\left(r\right)}^{\pi-\Theta\left(r\right)}P_{n}\left(\theta\right)\sin\theta d\theta\label{eq:swd-ell}\\
 & =\delta_{m0}\left\{ \sqrt{2}\delta_{n0}\mathcal{E}_{1}+\delta_{n,2k}\left(\mathcal{E}_{2}-\mathcal{E}_{1}\right)\dfrac{1}{\sqrt{4k+1}}\left[\dfrac{P_{2k+1}\left(\Theta\right)}{\sqrt{4k+3}}-\dfrac{P_{2k-1}\left(\Theta\right)}{\sqrt{4k-1}}\right]\right\} \nonumber 
\end{align}
Symbol $\mathcal{E}$ stands for any of functions in (\ref{eq:omega_dec}).
Due to the inner summation in Eqs. (\ref{eq:q_def}) decompositions
of the permittivity functions into the spherical harmonics should
be done for twice larger maximum harmonic degree $2N$ than the one
used in the methods. Functions (\ref{eq:q_matr}) are also simplified
since indices $m=q$ (for further simplification of the method based
on the rotational symmetry, see \cite{Doicu2018}).

In case of a general polyhedron particle the following approach can
be applied (see Fig. 4 for illustration). First, the integration over
the polar angle $\theta$ has to be replaced with the Gaussian quadrature
summation characterized by weights $w_{j}$ and angles $\theta_{j}$,
$j=1,\dots,N_{\theta}$. Note that a uniform series expansion for
an iteration-free calculation of the Gauss-Legendre nodes and weights
provided by \cite{Bogaert2014}, and used here, allows for an efficient
computation even of highly oscillating integrals at high precision.
Second, given a thin spherical shell of central radius $R_{k}$, intersections
of the sphere of the same radius with cones $\theta=\theta_{j}$ yield
a set of circles. In turn, being intersected with the polyhedron each
circle appears to be divided into an even number of arcs described
by azimuthal angles $\varphi_{jl}\left(R_{k}\right)$. Each of these
arcs lies either inside the polyhedron, and corresponds to some constant
$\mathcal{E}_{in}$, or outside the polyhedron, and corresponds to
another constant $\mathcal{E}_{out}$. Therefore, the integration
over the azimuthal angle can be done analytically, and the whole decomposition
becomes
\begin{align}
\left[\mathcal{E}\right]_{nm}\left(R_{k}\right) & \approx\frac{1}{2\pi}\sum_{j=0}^{N_{\theta}-1}w_{j}P_{n}^{m}\left(\theta_{j}\right)\sin\theta_{j}\left[\mathcal{E}_{in}\sum_{j}\intop_{\varphi_{jl}}^{\varphi_{jl+1}}\exp\left(-im\varphi\right)d\varphi+\mathcal{E}_{out}\sum_{j}\intop_{\varphi_{jl+1}}^{\varphi_{jl+2}}\exp\left(-im\varphi\right)d\varphi\right]\nonumber \\
 & =\sqrt{2}\mathcal{E}_{out}\delta_{m0}\delta_{n0}+\frac{\mathcal{E}_{in}-\mathcal{E}_{out}}{2\pi}\sum_{j=0}^{N_{\theta}-1}w_{j}P_{n}^{m}\left(\theta_{j}\right)\sin\theta_{j}\sum_{l}\Delta\varphi_{jl}\sinc\left(\frac{1}{2}m\Delta\varphi_{jl}\right)\exp\left[-\frac{1}{2}im\left(\varphi_{jl+1}+\varphi_{jl}\right)\right]\label{eq:swd-phd}
\end{align}
where $\Delta\varphi_{jl}=\varphi_{jl+1}-\varphi_{jl}$. For numerical
evaluation of Eq. (\ref{eq:swd-phd}) values of $P_{n}^{m}\left(\theta_{j}\right)$
can be pre-calculated, and then calculation of $\left[\mathcal{E}\right]_{nm}\left(R_{k}\right)$
can be done in parallel for a given set of shells.

\begin{figure}
\begin{centering}
\includegraphics[height=5cm]{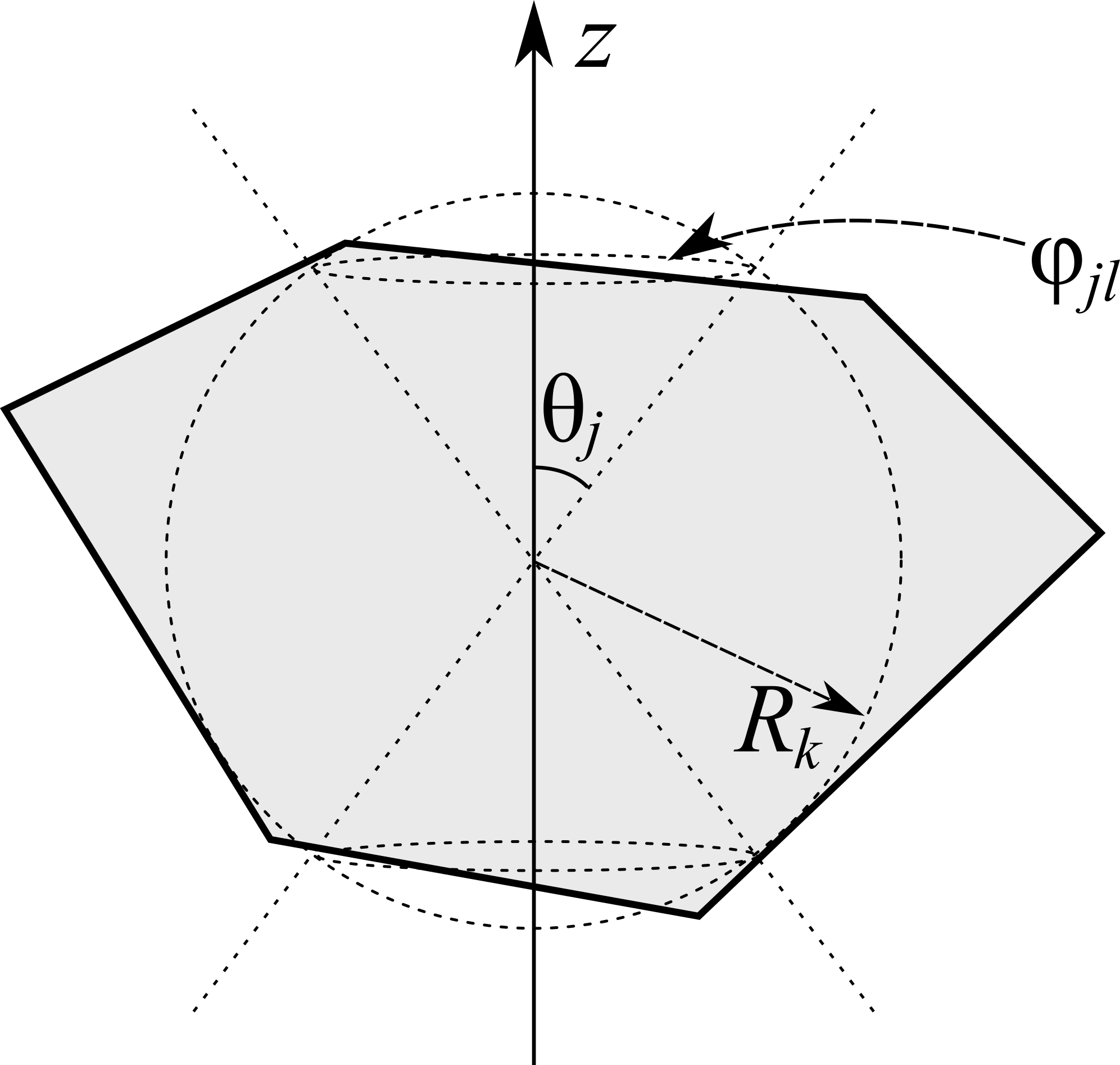}
\par\end{centering}
\caption{Illustration to the algorithm of the spherical harmonic transformation
of a polyhedron-sphere intersection.}

\end{figure}

When using either the scattering matrix or the scattering vector approaches
it is also preferable to benefit from the fact that solutions should
converge polynomially relative to increasing maximum degree of spherical
harmonics $N$ and the number of spherical shells $N_{s}$ used within
the discretization procedures. When such convergence is not significantly
perturbed by numerical errors, one can

\section{Numerical properties}

Accuracy of solutions of both the scattering matrix and the scattering
vector methods depends primarily on the number of spherical shells
$N_{s}$ and the maximum degree $N$ of spherical harmonic decomposition.
To attain accurate solutions of scattering problems the analysis is
done here in two steps. First, for fixed $N$ one looks for a limit
at increasing $N_{s}$, which corresponds to evaluation of integrals
in Eqs. (\ref{eq:ae_sol})-(\ref{eq:bh_sol}). Second, given the converged
solutions for several values of $N$ the convergence relative this
truncation number is traced. It was found that for fixed values of
$N$ convergence curves corresponding to increasing $N_{s}$ are all
pretty similar. Fig. 5 demonstrates an example of such convergence
for the scattering matrix method for a prolate spheroid of equatorial
size parameter $kd=10$, semi-axes ratio equal to $a/b=5$ and permittivity
$\varepsilon_{p}=4$. The vertical axis shows the maximum absolute
difference between scattering matrix components corresponding to subsequent
values of $N_{s}$. The rate of convergence is polynomial as follows
theoretically from the used integration rules, and this rate can be
substantially improved by the Richardson extrapolation. This is shown
by the two lower curves in Fig. 5 corresponding to the first and the
second order extrapolation. Fig. 6 demonstrates the accuracy of the
energy conservation law (power balance -- relation between incident
and scattered power minus 1) corresponding to the calculated and extrapolated
solutions of Fig. 5. The solutions converged relative to $N_{s}$
are then taken to trace the convergence for increasing $N$ (the accuracy
of radial integral evaluation should be enough not to affect this
latter convergence). Dependence of maximum absolute difference between
corresponding scattering matrix components for a prolate spheroid
of $kd=4$, $a/b=2$, and $\varepsilon_{p}=2$ is shown in Fig 7.

\begin{figure}
\begin{centering}
\includegraphics[height=7cm]{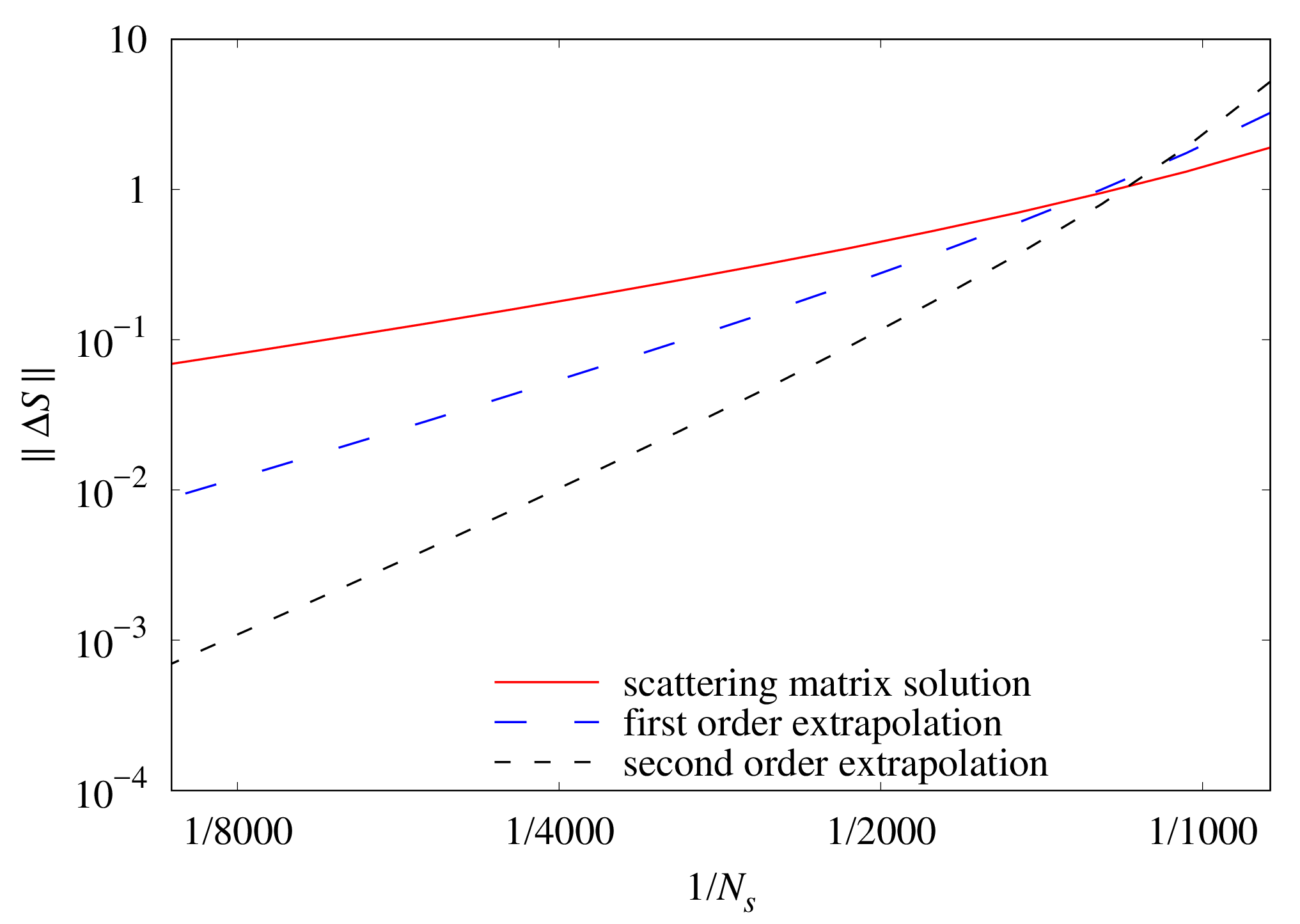}
\par\end{centering}
\caption{Convergence of the scattering matrix calculated by the scattering
matrix method for increasing number of spherical shells $N_{s}$ and
fixed spherical harmonic degree truncation number $N$ (the graphs
for different values of $N$ are pretty similar) in case of scattering
by a prolate spheroidal particle of equatorial size parameter $kd=10$,
semi-axes ratio $a/b=5$ and permittivity $\varepsilon_{p}=4$.}
\end{figure}

\begin{figure}
\begin{centering}
\includegraphics[height=7cm]{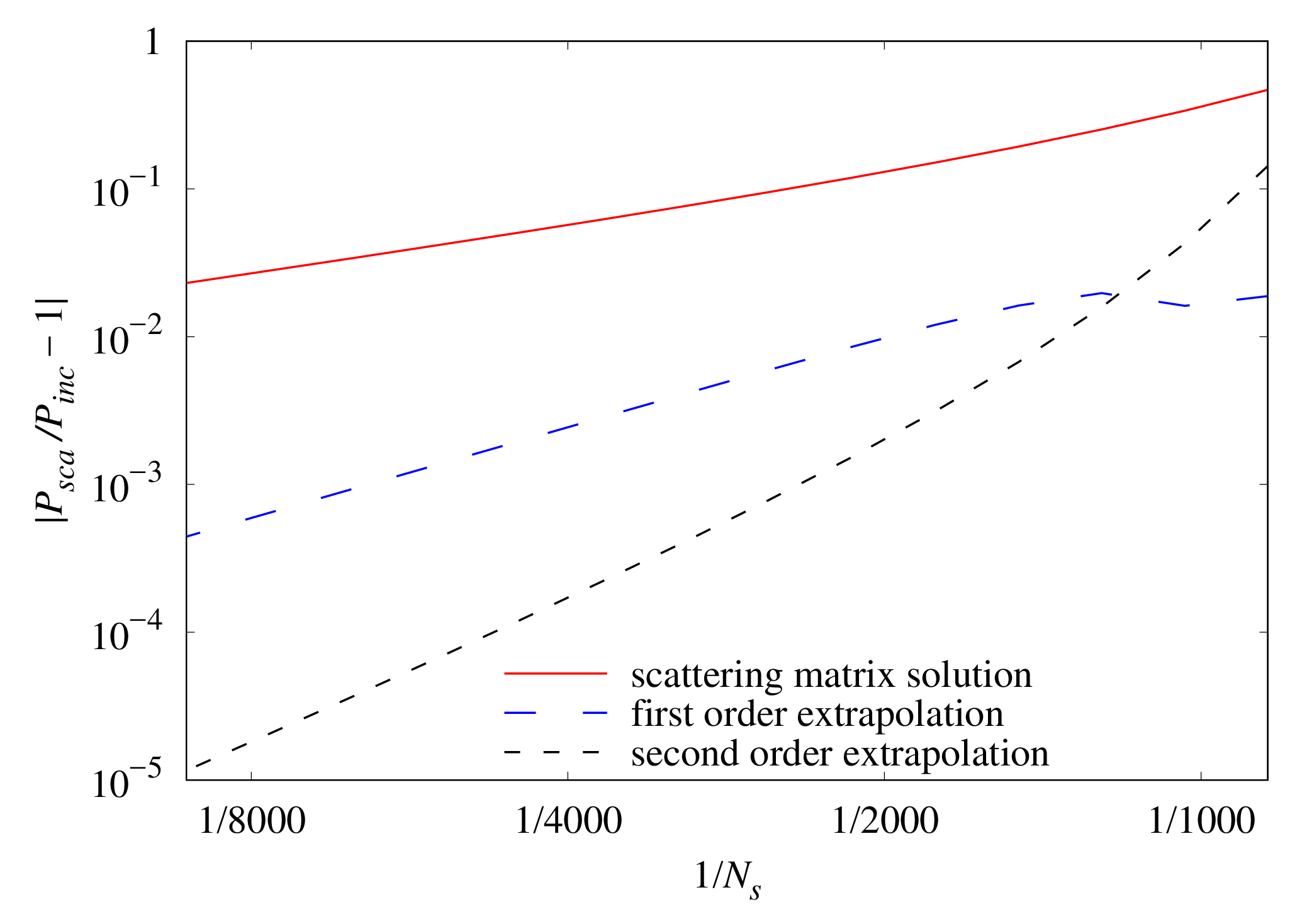}
\par\end{centering}
\caption{Power balance of the solutions attained by the scattering matrix method
and corresponding to the convergence shown in Fig. 5.}
\end{figure}

\begin{figure}
\begin{centering}
\includegraphics[height=7cm]{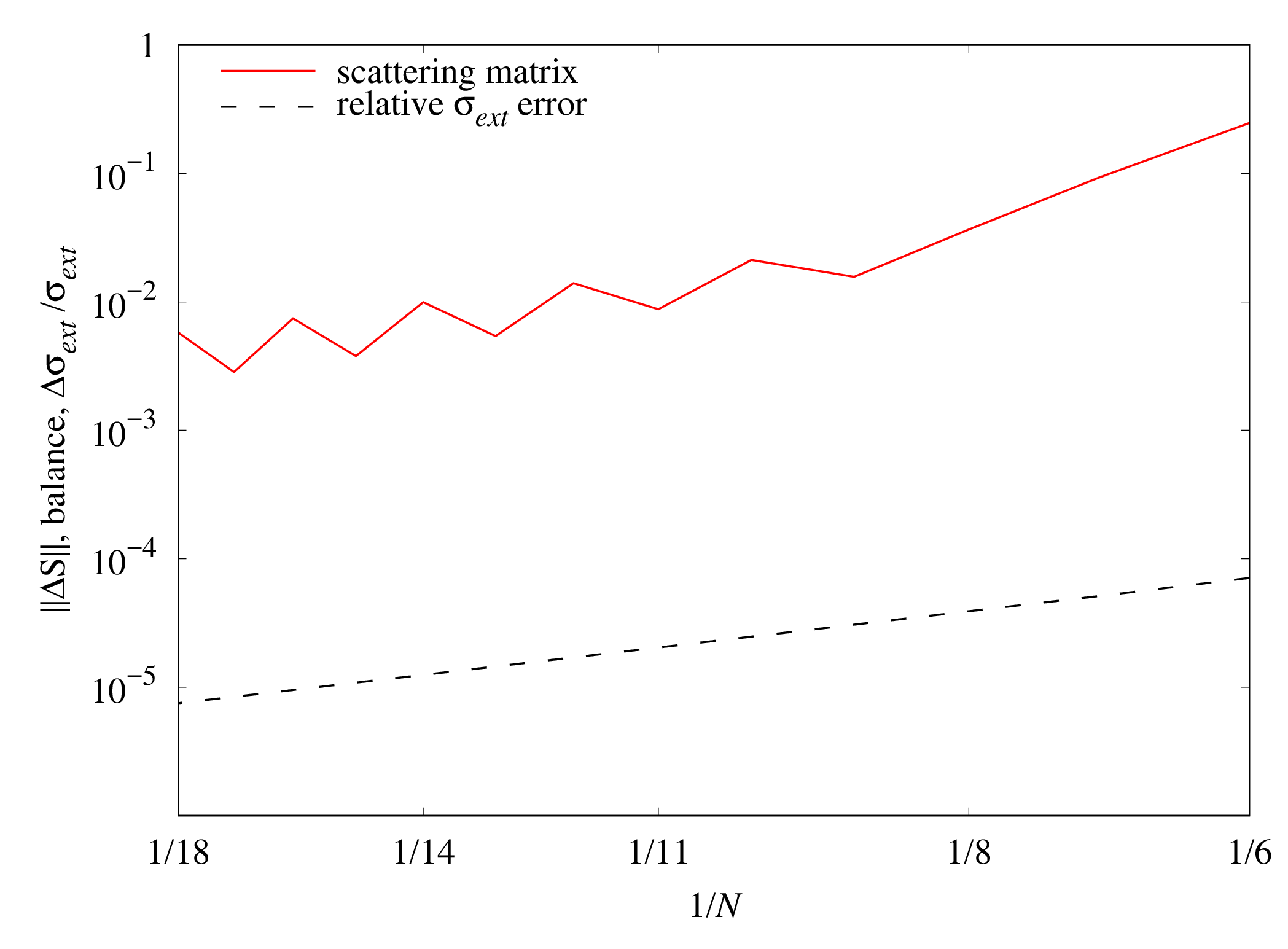}
\par\end{centering}
\caption{Convergence of the scattering matrix, and relative extinction cross
section calculated by the scattering matrix method for increasing
number of spherical harmonic degree truncation number $N$ providing
that each calculated matrix is converged over $N_{s}$ up to a sufficient
precision. This example is for scattering by prolate spheroidal particle
of equatorial size parameter $kd=4$, semi-axes ratio $a/b=2$ and
permittivity $\varepsilon_{p}=2$.}
\end{figure}

Figs. 8--10 demonstrate the same convergence plots as Figs. 5--7
for the scattering vector method based calculations. This method has
the second order polynomial convergence rate relative to increasing
$N_{s}$, so only the second order Richardson extrapolation is applied
here. Importantly, one can also notice that calculated scattering
vector solutions have almost zero power balance contrary to the scattering
matrix method, Fig. 6.

An example of a complex shape polyhedral particles was generated by
randomly shifting and stretching of icosahedron vertex positions.
Fig. 11 demonstrates an example of such a particle, which has no symmetries,
and Fig. 12 shows the convergence of the scattering vector method
applied to such particle with characteristic size (circumscribed sphere
diameter) $kD=8$ and permittivity 2.

Free parameter $\varepsilon_{b}$ fairly affects the results of the
scattering matrix method when being chosen to be pure real and to
vary within the interval $\varepsilon_{s}\leq\varepsilon_{b}\leq\text{\ensuremath{\Re\varepsilon_{p}}}$.
On the opposite, this parameter may slightly affect the convergence
speed of a linear solver of the equation system (\ref{eq:svector_solution})
being for example the GMRes or the BiCGstab. This convergence was
also found not only to depend predictably on the refractive index
contrast and the size parameter, but to increase with the increasing
number $N$, which may probably be related to a loss of accuracy.
Viz, for a given scattering particle the number of iterations $n_{it}$
required to solve the system (\ref{eq:svector_solution}) is independent
of $N$ for relatively small $N$ but at some point starts to substantially
increase. Amid this effect the stagnation of the iterative process
occurs. To overcome these barriers and to gain the most of the parallelization
potential of the scattering vector algorithm it seems to be essential
to search for a suitable preconditioner which is a subject of a future
research.

\begin{figure}
\begin{centering}
\includegraphics[height=7cm]{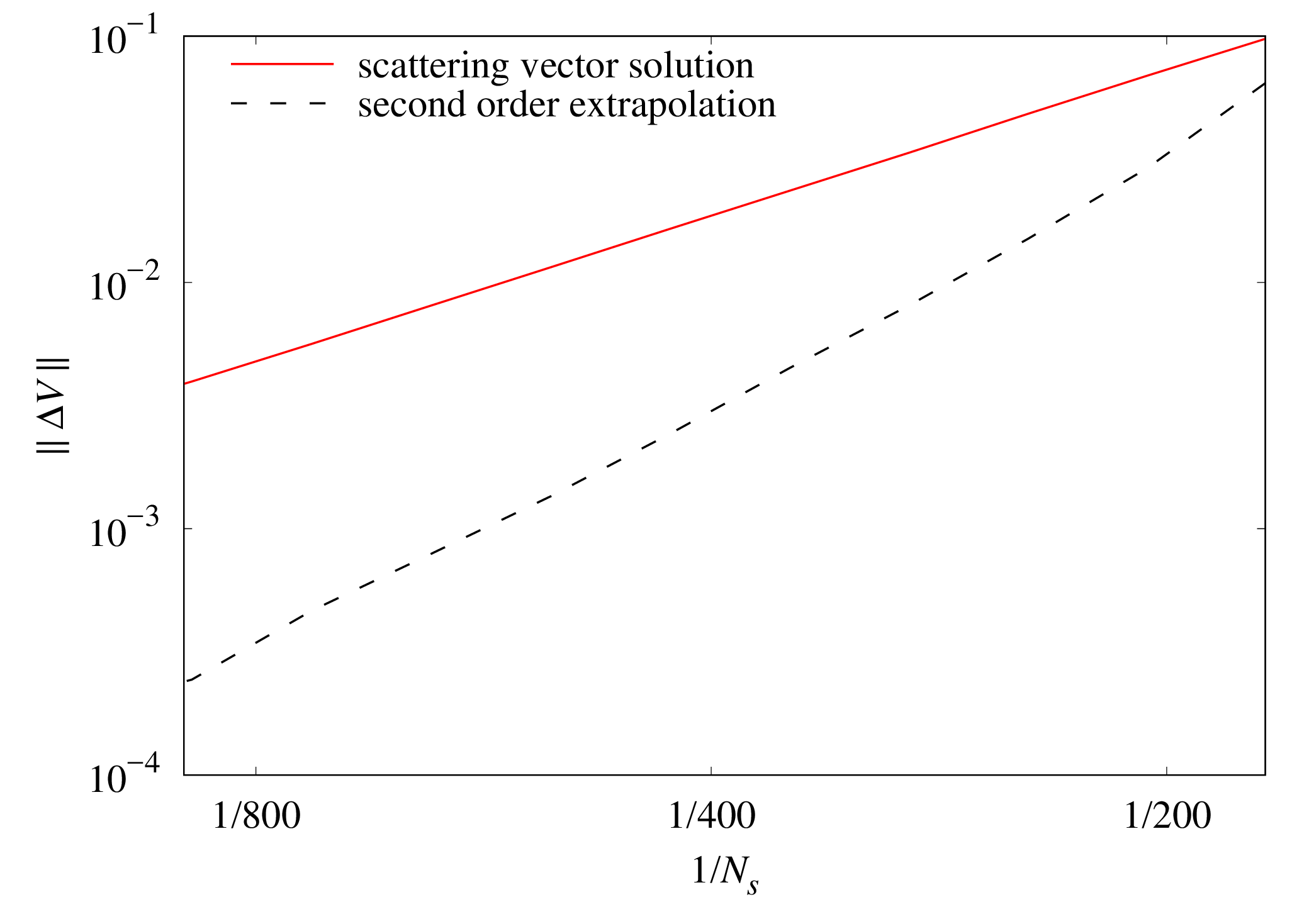}
\par\end{centering}
\caption{Same as in Fig. 5, but for the scattering vector method.}
\end{figure}

\begin{figure}
\begin{centering}
\includegraphics[height=7cm]{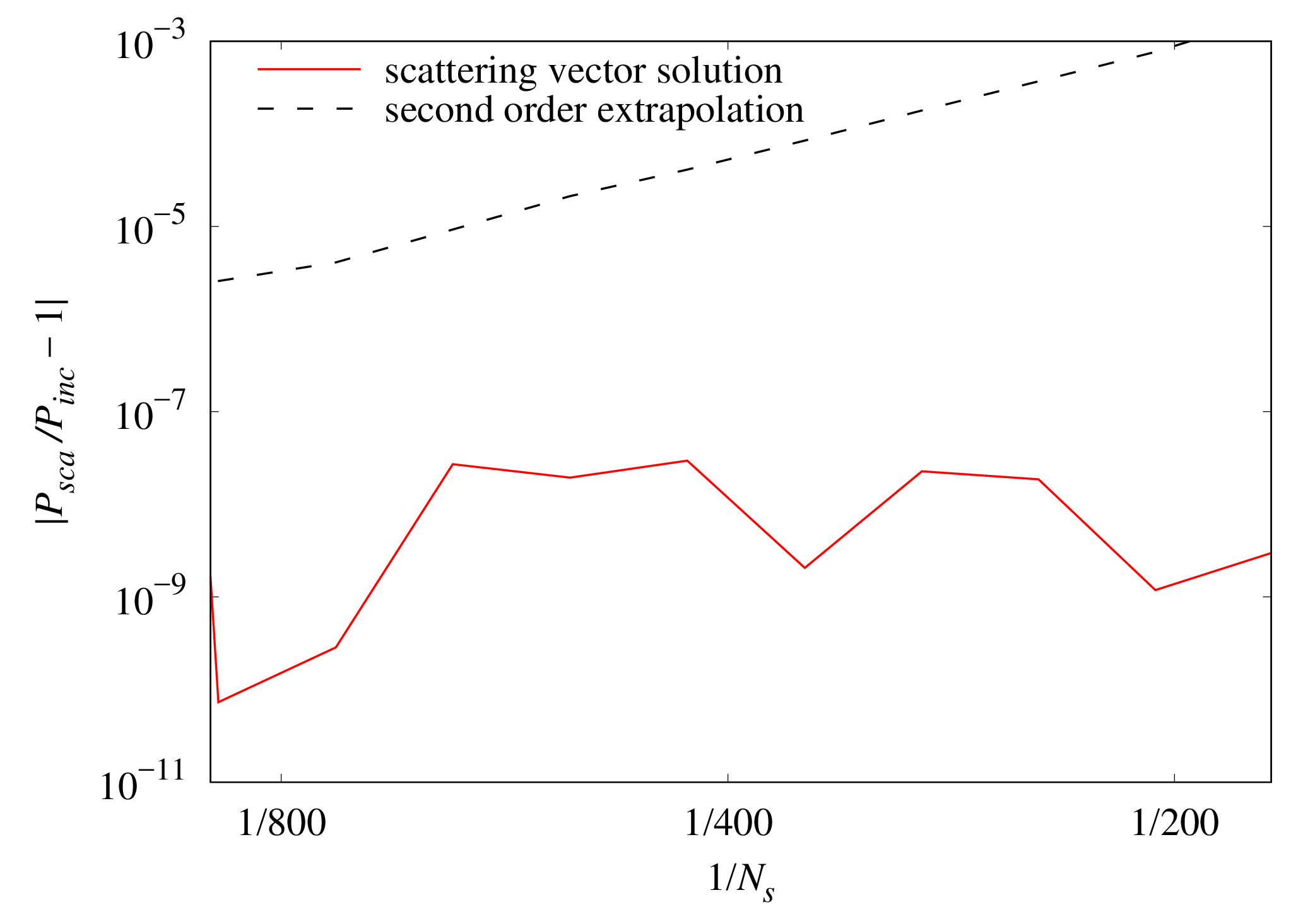}
\par\end{centering}
\caption{Same as in Fig. 6, but for the scattering vector method.}
\end{figure}

\begin{figure}
\begin{centering}
\includegraphics[height=7cm]{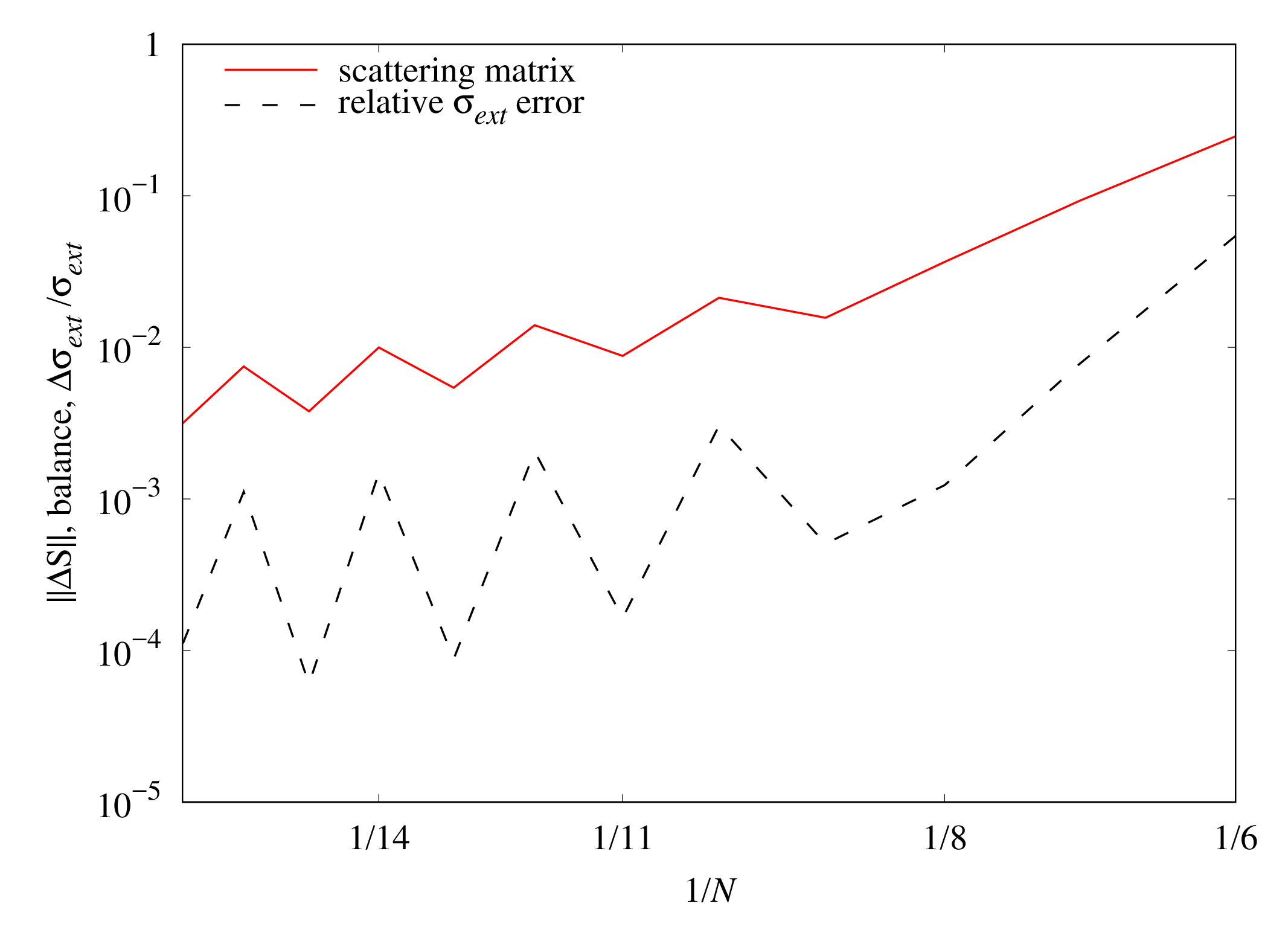}
\par\end{centering}
\caption{Same as in Fig. 7, but for the scattering vector method.}

\end{figure}

\begin{figure}
\begin{centering}
\includegraphics[height=5cm]{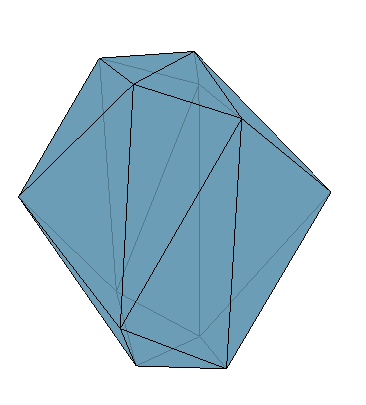}
\par\end{centering}
\caption{Example of a scattering particle of irregular shape. The particle
is generated by randomly shifting positions of vertices of a regular
icosahedron.}
\end{figure}

\begin{figure}
\begin{centering}
\includegraphics[height=7cm]{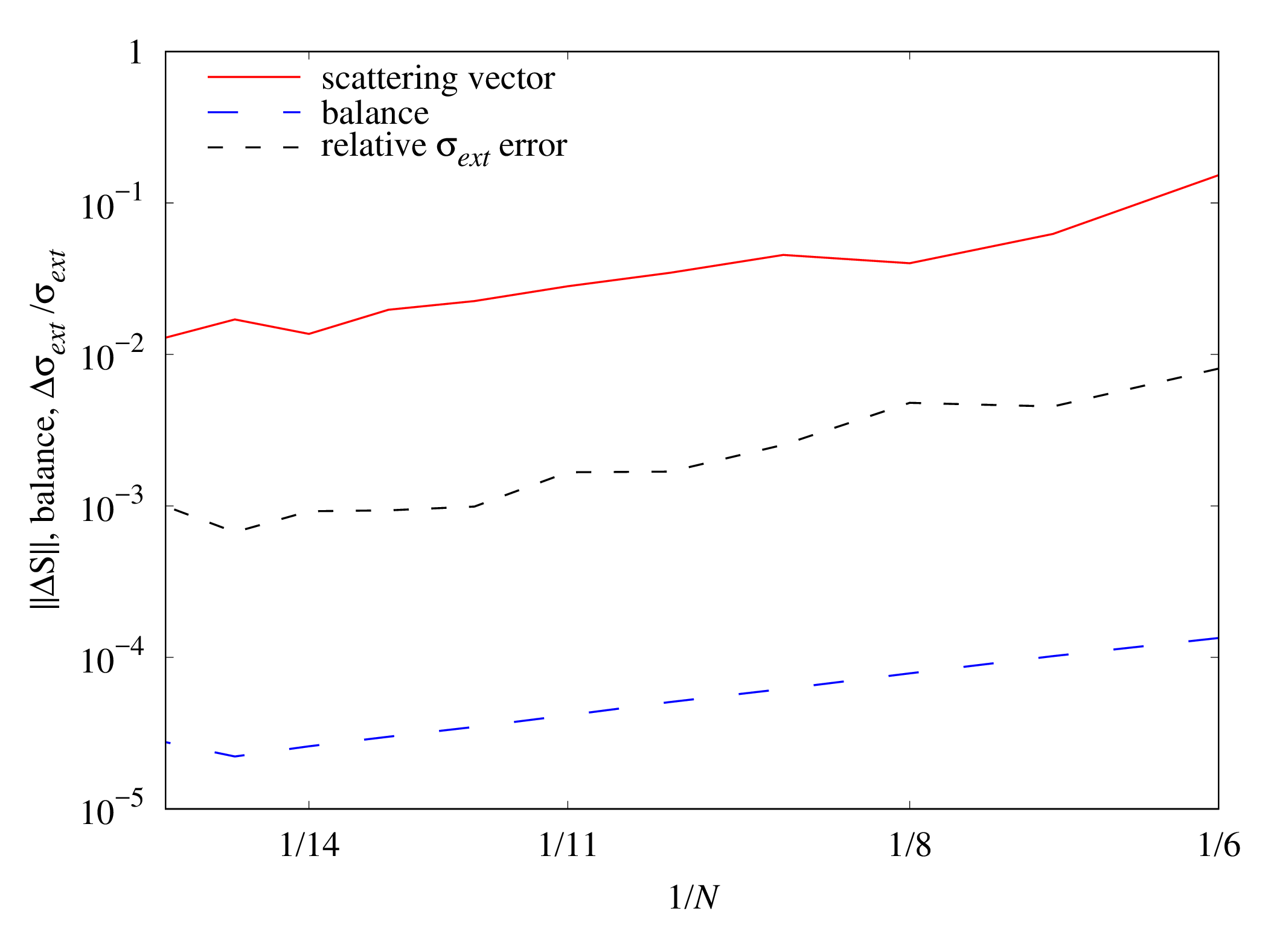}
\par\end{centering}
\caption{Convergence of the scattering matrix, power balance and the relative
extinction cross section calculated by the scattering matrix method
for increasing number of spherical harmonic degree truncation number
$N$ providing that each calculated matrix is converged over $N_{s}$
up to a sufficient precision. This example is for scattering by an
irregular polyhedral particle of size parameter $kd=8$, and permittivity
$\varepsilon_{p}=2$.}

\end{figure}

\section{Summary}

To conclude, the work provides the derivation of a method analogous
to the IIM on the basis of the Generalized Source approach. In analogy
with the planar geometry case the method is referred to as the scattering
matrix method. The Generalized Source viewpoint will allow to extend
the approach to the curvilinear coordinate transformations and related
metric sources in a next publication. Simultaneously an alternative
to the scattering matrix method, the method of calculating separate
scattering matrix columns -- the scattering vector method, is developed.
The latter method relies on the iterative solution of linear equation
systems and adopts parallelization which makes it potentially attractive
for large scale computations, though an additional work to improve
its numerical behavior is required. It is shown that the scattering
vector method yields solutions with almost zero power balance when
applied to dielectric particles. In addition the paper proposes an
algorithm of computing of spherical harmonic transformations for cross-sections
coming from crossing of arbitrary polyhedral shape scattering particles
by thin spherical shells.

\section*{Acknowledgments}

The work was supported by the Russian Science Foundation, grant no.
17-79-20418.

\section*{Appendix A}

This appendix lists the explicit components of the scattering matrix
present in Eq. (\ref{eq:S-mat})
\begin{equation}
S_{nm,pq}^{11,ee}\left(x,\Delta x\right)=i\Delta x{\bf \mathcal{V}}_{n1}^{(1)}\left(x\right)Q_{nm,pq}^{+}\left(x\right){\bf \mathcal{V}}_{p1}^{(1)}\left(x\right)
\end{equation}
\begin{equation}
S_{nm,pq}^{11,eh}\left(x,\Delta x\right)=i\Delta x{\bf \mathcal{V}}_{n1}^{(1)}\left(x\right)Q_{nm,pq}^{-}\left(x\right){\bf \mathcal{V}}_{p2}^{(1)}\left(x\right)
\end{equation}
\begin{equation}
S_{nm,pq}^{11,he}\left(x,\Delta x\right)=-i\Delta x{\bf \mathcal{V}}_{n2}^{(1)}\left(x\right)Q_{nm,pq}^{-}\left(x\right){\bf \mathcal{V}}_{p1}^{(1)}\left(x\right)
\end{equation}
\begin{equation}
S_{nm,pq}^{11,hh}\left(x,\Delta x\right)=i\Delta x\left[{\bf \mathcal{V}}_{n3}^{(1)}\left(x\right)Q_{nm,pq}^{r}\left(x\right){\bf \mathcal{V}}_{p3}^{(1)}\left(x\right)-{\bf \mathcal{V}}_{n2}^{(1)}\left(x\right)Q_{nm,pq}^{+}\left(x\right){\bf \mathcal{V}}_{p2}^{(1)}\left(x\right)\right]
\end{equation}
\begin{equation}
S_{nm,pq}^{12,ee}\left(x,\Delta x\right)=1+i\Delta x{\bf \mathcal{V}}_{n1}^{(1)}\left(x\right)Q_{nm,pq}^{+}\left(x\right){\bf \mathcal{V}}_{p1}^{(3)}\left(x\right)
\end{equation}
\begin{equation}
S_{nm,pq}^{12,eh}\left(x,\Delta x\right)=i\Delta x{\bf \mathcal{V}}_{n1}^{(1)}\left(x\right)Q_{nm,pq}^{-}\left(x\right){\bf \mathcal{V}}_{p2}^{(1)}\left(x\right)
\end{equation}
\begin{equation}
S_{nm,pq}^{12,he}\left(x,\Delta x\right)=-i\Delta x{\bf \mathcal{V}}_{n2}^{(1)}\left(x\right)Q_{nm,pq}^{-}\left(x\right){\bf \mathcal{V}}_{p1}^{(3)}\left(x\right)
\end{equation}
\begin{equation}
S_{nm,pq}^{12,hh}\left(x,\Delta x\right)=1+i\Delta x\left[{\bf \mathcal{V}}_{n3}^{(1)}\left(x\right)Q_{nm,pq}^{r}\left(x\right){\bf \mathcal{V}}_{p3}^{(3)}\left(x\right)-{\bf \mathcal{V}}_{n2}^{(1)}\left(x\right)Q_{nm,pq}^{+}\left(x\right){\bf \mathcal{V}}_{p2}^{(3)}\left(x\right)\right]
\end{equation}
\begin{equation}
S_{nm,pq}^{21,ee}\left(x,\Delta x\right)=1+i\Delta x{\bf \mathcal{V}}_{n1}^{(3)}\left(x\right)Q_{nm,pq}^{+}\left(x\right){\bf \mathcal{V}}_{p1}^{(1)}\left(x\right)
\end{equation}
\begin{equation}
S_{nm,pq}^{21,eh}\left(x,\Delta x\right)=i\Delta x{\bf \mathcal{V}}_{n1}^{(3)}\left(x\right)Q_{nm,pq}^{-}\left(x\right){\bf \mathcal{V}}_{p2}^{(1)}\left(x\right)
\end{equation}
\begin{equation}
S_{nm,pq}^{21,he}\left(x,\Delta x\right)=-i\Delta x{\bf \mathcal{V}}_{n2}^{(3)}\left(x\right)Q_{nm,pq}^{-}\left(x\right){\bf \mathcal{V}}_{p1}^{(1)}\left(x\right)
\end{equation}
\begin{equation}
S_{nm,pq}^{21,hh}\left(x,\Delta x\right)=1+i\Delta x\left[{\bf \mathcal{V}}_{n3}^{(3)}\left(x\right)Q_{nm,pq}^{r}\left(x\right){\bf \mathcal{V}}_{p3}^{(1)}\left(x\right)-{\bf \mathcal{V}}_{n2}^{(3)}\left(x\right)Q_{nm,pq}^{+}\left(x\right){\bf \mathcal{V}}_{p2}^{(1)}\left(x\right)\right]
\end{equation}
\begin{equation}
S_{nm,pq}^{22,ee}\left(x,\Delta x\right)=i\Delta x{\bf \mathcal{V}}_{n1}^{(3)}\left(x\right)Q_{nm,pq}^{+}\left(x\right){\bf \mathcal{V}}_{p1}^{(3)}\left(x\right)
\end{equation}
\begin{equation}
S_{nm,pq}^{22,eh}\left(x,\Delta x\right)=i\Delta x{\bf \mathcal{V}}_{n1}^{(3)}\left(x\right)Q_{nm,pq}^{-}\left(x\right){\bf \mathcal{V}}_{p2}^{(1)}\left(x\right)
\end{equation}
\begin{equation}
S_{nm,pq}^{22,he}\left(x,\Delta x\right)=-i\Delta x{\bf \mathcal{V}}_{n2}^{(3)}\left(x\right)Q_{nm,pq}^{-}\left(x\right){\bf \mathcal{V}}_{p1}^{(3)}\left(x\right)
\end{equation}

\begin{equation}
S_{nm,pq}^{22,hh}\left(x,\Delta x\right)=i\Delta x\left[{\bf \mathcal{V}}_{n3}^{(3)}\left(x\right)Q_{nm,pq}^{r}\left(x\right){\bf \mathcal{V}}_{p3}^{(3)}\left(x\right)-{\bf \mathcal{V}}_{n2}^{(3)}\left(x\right)Q_{nm,pq}^{+}\left(x\right){\bf \mathcal{V}}_{p2}^{(3)}\left(x\right)\right]
\end{equation}

Multiplication of the amplitude vector $\left(\tilde{a}_{nm}^{e}\left(x\right),\thinspace\tilde{b}_{nm}^{e}\left(x\right),\thinspace\tilde{a}_{nm}^{h}\left(x\right),\thinspace\tilde{b}_{nm}^{h}\left(x\right)\right)^{T}$
by this matrix can be performed in the following three steps:
\begin{enumerate}
\item Calculate intermediate coefficient vectors
\begin{equation}
\begin{array}{c}
\xi_{nm}^{1}\left(x\right)={\bf \mathcal{V}}_{n1}^{(3)}\left(x\right)\tilde{a}_{nm}^{e}\left(x\right)+{\bf \mathcal{V}}_{n1}^{(1)}\left(x\right)\tilde{b}_{nm}^{e}\left(x\right)\\
\xi_{nm}^{2}\left(x\right)={\bf \mathcal{V}}_{n2}^{(3)}\left(x\right)\tilde{a}_{nm}^{h}\left(x\right)+{\bf \mathcal{V}}_{n2}^{(1)}\left(x\right)\tilde{b}_{nm}^{h}\left(x\right)\\
\xi_{nm}^{3}\left(x\right)={\bf \mathcal{V}}_{n3}^{(3)}\left(x\right)\tilde{a}_{nm}^{h}\left(x\right)+{\bf \mathcal{V}}_{n3}^{(1)}\left(x\right)\tilde{b}_{nm}^{h}\left(x\right)
\end{array}
\end{equation}
\item Perform matrix-vector multiplications
\begin{equation}
\begin{array}{c}
\Xi_{nm}^{1}\left(x\right)=\sum_{pq}\left[Q_{nm;pq}^{+}\left(x\right)\xi_{pq}^{1}\left(x\right)+Q_{nm;pq}^{-}\left(x\right)\xi_{pq}^{2}\left(x\right)\right]\\
\Xi_{nm}^{2}\left(x\right)=\sum_{pq}\left[Q_{nm;pq}^{+}\left(x\right)\xi_{pq}^{2}\left(x\right)+Q_{nm;pq}^{-}\left(x\right)\xi_{pq}^{1}\left(x\right)\right]\\
\Xi_{nm}^{3}\left(x\right)=\sum_{pq}Q_{nm;pq}^{r}\left(x\right)\xi_{pq}^{3}\left(x\right)
\end{array}
\end{equation}
\item Find scattered field amplitudes
\begin{equation}
\begin{array}{c}
\tilde{a}_{nm}^{sca,e}\left(x\right)=\tilde{a}_{nm}^{e}\left(x\right)+i\Delta x{\bf \mathcal{V}}_{n1}^{(1)}\left(x\right)\Xi_{nm}^{1}\left(x\right)\\
\tilde{b}_{nm}^{sca,e}\left(x\right)=\tilde{b}_{nm}^{e}\left(x\right)+i\Delta x{\bf \mathcal{V}}_{n1}^{(3)}\left(x\right)\Xi_{nm}^{1}\left(x\right)\\
\tilde{a}_{nm}^{sca,h}\left(x\right)=\tilde{a}_{nm}^{h}\left(x\right)+i\Delta x\left[{\bf \mathcal{V}}_{n3}^{(1)}\Xi_{nm}^{3}\left(x\right)-{\bf \mathcal{V}}_{n2}^{(1)}\Xi_{nm}^{2}\left(x\right)\right]\\
\tilde{b}_{nm}^{sca,h}\left(x\right)=\tilde{b}_{nm}^{h}\left(x\right)+i\Delta x\left[{\bf \mathcal{V}}_{n3}^{(3)}\Xi_{nm}^{3}\left(x\right)-{\bf \mathcal{V}}_{n2}^{(3)}\Xi_{nm}^{2}\left(x\right)\right]
\end{array}
\end{equation}
\end{enumerate}

\section*{Appendix B}

This Appendix provides an explicit scattering matrix of the outer
spherical interface separating the basis and the surrounding media
having permittivities $\varepsilon_{b}$ and $\varepsilon_{s}$ respectively.
Let us consider the field decomposition into the vector spherical
waves in the vicinity of the outer spherical interface $r=R_{out}$.
For $r=R_{out}-0$ the field is represented as a superposition of
regular and outgoing spherical wave functions, whereas for $r=R_{out}+0$
-- as a superposition of incoming and outgoing waves:
\begin{equation}
{\bf E}\left(r,\theta,\varphi\right)=\begin{cases}
\sum_{nm}a_{nm}^{(b)e}\mathcal{{\bf \mathcal{M}}}_{nm}^{3}\left(k_{b}\bm{r}\right)+a_{nm}^{(b)h}\mathcal{{\bf \mathcal{N}}}_{nm}^{3}\left(k_{b}\bm{r}\right)+b_{nm}^{(b)e}\mathcal{{\bf \mathcal{M}}}_{nm}^{1}\left(k_{b}\bm{r}\right)+b_{nm}^{(b)h}\mathcal{{\bf \mathcal{N}}}_{nm}^{1}\left(k_{b}\bm{r}\right) & r=R_{out}-0\\
\sum_{nm}a_{nm}^{(s)e}\mathcal{{\bf \mathcal{M}}}_{nm}^{3}\left(k_{s}\bm{r}\right)+a_{nm}^{(s)h}\mathcal{{\bf \mathcal{N}}}_{nm}^{3}\left(k_{s}\bm{r}\right)+c_{nm}^{(s)e}\mathcal{{\bf \mathcal{M}}}_{nm}^{2}\left(k_{s}\bm{r}\right)+c_{nm}^{(s)h}\mathcal{{\bf \mathcal{N}}}_{nm}^{2}\left(k_{s}\bm{r}\right) & r=R_{out}+0
\end{cases}
\end{equation}
The magnetic field is directly obtained from the Faraday's law, first
of Eq. (\ref{eq:maxwell}), and transformation rules Eq. (\ref{eq:MN_NM}).
Relation between coefficients in the field decomposition comes from
the boundary condition at the interface $r=R_{out}$ consisting in
continuity of the tangential field components $E_{\pm}\left(R_{out}-0,\theta,\varphi\right)=E_{\pm}\left(R_{out}+0,\theta,\varphi\right)$,
and $H_{\pm}\left(R_{out}-0,\theta,\varphi\right)=H_{\pm}\left(R_{out}+0,\theta,\varphi\right)$.
The orthogonality relations for $\mathcal{{\bf \mathcal{M}}}_{nm}^{1,3}$,
and $\mathcal{{\bf \mathcal{N}}}_{nm}^{1,3}$ \cite{Doicu2006} together
with spherical Bessel function Wronskian relations bring the following
explicit formulas in form of scattering matrix transformations:
\begin{align}
\left(\begin{array}{c}
b_{nm}^{(b)e}\\
a_{nm}^{(s)e}
\end{array}\right) & =\dfrac{1}{j_{n}\left(k_{b}R_{out}\right)\tilde{h}_{n}^{(1)}\left(k_{s}R_{out}\right)-h_{n}^{(1)}\left(k_{s}R_{out}\right)\tilde{j}_{n}\left(k_{b}R_{out}\right)}\nonumber \\
 & \times\left(\begin{array}{cc}
\left[\begin{array}{c}
h_{n}^{(1)}\left(k_{s}R_{out}\right)\tilde{h}_{n}^{(1)}\left(k_{b}R_{out}\right)\\
-h_{n}^{(1)}\left(k_{b}R_{out}\right)\tilde{h}_{n}^{(1)}\left(k_{s}R_{out}\right)
\end{array}\right] & \dfrac{2i}{k_{s}R_{out}}\\
\dfrac{i}{k_{b}R_{out}} & \left[\begin{array}{c}
h_{n}^{(2)}\left(k_{s}R_{out}\right)\tilde{j}_{n}\left(k_{b}R_{out}\right)\\
-j_{n}\left(k_{b}R_{out}\right)\tilde{h}_{n}^{(2)}\left(k_{s}R_{out}\right)
\end{array}\right]
\end{array}\right)\left(\begin{array}{c}
a_{nm}^{(b)e}\\
c_{nm}^{(s)e}
\end{array}\right)
\end{align}
\begin{align}
\left(\begin{array}{c}
b_{nm}^{(b)h}\\
a_{nm}^{(s)h}
\end{array}\right) & =\dfrac{1}{\dfrac{\varepsilon_{b}}{\varepsilon_{s}}j_{n}\left(k_{b}R_{out}\right)\tilde{h}_{n}^{(1)}\left(k_{s}R_{out}\right)-h_{n}^{(1)}\left(k_{s}R_{out}\right)\tilde{j}_{n}\left(k_{b}R_{out}\right)}\nonumber \\
 & \times\left(\begin{array}{cc}
\left[\begin{array}{c}
h_{n}^{(1)}\left(k_{s}R_{out}\right)\tilde{h}_{n}^{(1)}\left(k_{b}R_{out}\right)\\
-\dfrac{\varepsilon_{b}}{\varepsilon_{s}}h_{n}^{(1)}\left(k_{b}R_{out}\right)\tilde{h}_{n}^{(1)}\left(k_{s}R_{out}\right)
\end{array}\right] & \sqrt{\dfrac{\varepsilon_{b}}{\varepsilon_{s}}}\dfrac{2i}{k_{s}R_{out}}\\
\dfrac{i}{k_{s}R_{out}} & \left[\begin{array}{c}
h_{n}^{(2)}\left(k_{s}R_{out}\right)\tilde{j}_{n}\left(k_{b}R_{out}\right)\\
-\dfrac{\varepsilon_{b}}{\varepsilon_{s}}j_{n}\left(k_{b}R_{out}\right)\tilde{h}_{n}^{(2)}\left(k_{s}R_{out}\right)
\end{array}\right]
\end{array}\right)\left(\begin{array}{c}
a_{nm}^{(b)h}\\
c_{nm}^{(s)h}
\end{array}\right)
\end{align}

\bibliographystyle{IEEEtran}
\bibliography{sphere}

\end{document}